\documentstyle[prb,preprint,aps,epsf]{revtex}
\draft

\def\AmS{{\protect\the\textfont2
        A\kern-.1667em\lower.5ex\hbox{M}\kern-.125emS}}

\makeatletter
\def\thepage{1-\@arabic\c@page}
\def\@pnumwidth{2em}
\makeatother
\begin{document}

\title{Two Dimensional Quantum Mechanical Modeling of Nanotransistors}

\author{A. Svizhenko, M. P. Anantram,\cite{byline} T. R. Govindan, B. Biegel}
\address{
NASA Ames Research Center, 
Mail Stop: T27A-1, 
Moffett Field, 
CA 94035-1000, U.S.A.
}

\author{R. Venugopal}
\address{
  School of Electrical and Computer Engineering,
  Purdue University, West Lafayette, IN 47907-1285
}

\maketitle

\vspace{0.3in}

\centerline{\bf Accepted for publication in the Journal of Applied Physics}

\begin{abstract}
Quantization in the inversion layer and phase coherent transport
are anticipated to have significant impact on device performance
in 'ballistic' nanoscale transistors. While the role of some quantum 
effects have been analyzed qualitatively using simple one dimensional
ballistic models, two dimensional (2D) quantum mechanical simulation is
important for quantitative results. In this paper, we present a 
framework for 2D quantum mechanical simulation of a nanotransistor /
Metal Oxide Field Effect Transistor (MOSFET). This 
framework consists of the non equilibrium Green's function
equations solved self-consistently with Poisson's equation. Solution of
this set of equations is computationally intensive. An efficient 
algorithm to calculate the quantum mechanical 2D electron density 
has been developed. The method presented is comprehensive in that 
treatment includes the three open boundary conditions, where the narrow
channel region opens into physically broad source, drain and gate 
regions. Results are presented for 
(i) drain current versus drain and gate voltages,
(ii) comparison to results from Medici, and (iii) gate tunneling current,
using 2D potential profiles. 
Methods to reduce the gate leakage current are also discussed based on
simulation results.

\end{abstract}
\pacs{ }

\narrowtext

\section{Introduction}
\label{sec:introduction}
MOSFETs with channel lengths in the tens of nanometer regime have
recently been demonstrated by various research 
labs.~\cite{Hergenrother99,Timp99,Appenzeller00} Design considerations
to yield devices with desirable characteristics have been explored in
references \onlinecite{Taur97,Taur98,Wong99,Thompson,SIA97}.
Device physics of these MOSFETs were analysed using simple quasi one
dimensional models.~\cite{Natori94,Lundstrom97,Ren00,Naveh00,Ren00_2}
The best modeling approach for design and analysis of nanoscale MOSFETs
is presently unclear,
though a straightforward application of semiclassical methods that
disregards quantum mechanical effects is generally accepted to be
inadequate. Quantum mechanical modeling of MOSFETs with channel lengths
in the tens of nanometers is important for many reasons:

\noindent
(i) MOSFETs with ultrathin oxide require an accurate treatment of
current injection from source, drain and gate. Gate leakage is
important because it places a lower limit on the OFF current.

\noindent
(ii) Ballistic flow of electrons across the channel becomes 
increasingly important as the channel length decreases.

\noindent
(iii) The location of the inversion layer changes from the source to
the drain end, and its role in determining the C-V and I-V
characterestics is most accurately included by a self-consistent
solution of Poisson's equation and a quantum mechanical description to
compute the charge density.

\noindent
(iv) Approximate theories of quantum effects included in semi-classical
MOSFET modeling tools are desirable from practical considerations
because semi-classical methods are numerically less expensive, and much
of the empirical and semi-classical MOSFET physics developed over the
last few
decades continues to hold true in many regions of a nanoscale MOSFET.
Examples of semiclassical methods that consider some quantum mechanical
aspects are the density gradient,~\cite{Biegel98,Ancona87} and effective
potential~\cite{Ferry00} approaches, and quantum mechanical
approximations used in the Medici package.~\cite{Medici}
Fully quantum mechanical simulations can play an important role in
benchmarking such simulators.

Central to quantum mechanical approaches to charge transport modeling
is self-consistent solution of a wave equation to describe the quantum
mechanical transport, Poisson's equation, and equations for statistics
of the  particle ensemble. In the absence of electron-electron
and electron-phonon interactions (state of the scatterer does not
change), the Landauer-Buttiker formalism~\cite{Buttiker93,Datta_book}
is applicable. In this
formalism, the wave equation is Schrodinger's equation and the
statistics is represented throughout the device by the Fermi-Dirac
distribution of particles incident from the contacts (source, drain and
gate). In the presence of electron-phonon
interaction, the Wigner function (WF) and non equilibrium Green's
function (NEGF) formalisms are used. The NEGF approach has been quite
successful in modeling steady state transport in a wide variety of one
dimensional (1D) semiconductor structures.~\cite{Lake97,NEMO}

A number of groups have started developing theory and simulation for fully
quantum mechanical {\it two dimensional} simulation of MOSFETs using the:
real space approach,~\cite{Abramo97,Fu97,Spinelli98} k-space
approach,~\cite{Abramo00}  Wigner function approach,~\cite{Han00} and 
non equilibrium Green's function approach.~\cite{Ren00_2,Jovanovic00,Svizhenko00}
Others groups have taken a hybrid approach
using the Monte Carlo method. The Monte Carlo approach, has been quite
successful in describing scattering mechanisms in MOSFETs, in 
comparison to fully quantum mechanical approaches, and can also include
ballistic effects and the role of quantized energy levels in the MOSFET
inversion layer in an approximate 
manner.~\cite{Fischetti93,Ravaioli00,Vasileska00} Discussing the 
relative merits of various approaches and quantum-corrected 
drift-diffusion approaches is important. In fact, such a comparison of
methods using standard device structures has been initiated~\cite{NASA3}
but much work remains to be done in comparing and studying the
suitability of different methods. Comparison of various methods is {\it
not} the purpose of this paper.
The purpose of this paper is to describe development of a particular
approach, namely the NEGF approach, for numerical simulation of MOSFETs
with two dimensional (2D) doping profiles and charge injection 
from the source, drain and gate contacts. 2D simulation significantly
increases computational effort over the 1D case.
Non-uniform spatial grids are essential to limit the total number of
grid points while at the same time resolving physical features.
A new algorithm for efficient computation of electron density without
complete solution of the system of equations is presented. The computer
code developed is used to calculate the drain and gate tunneling 
current in ultra short channel MOSFETs. Results from our approach and
Medici are compared. The paper is organized as follows:
formalism (section \ref{section:Formalism}), role of
polysilicon gate depletion \ref{subsection:poly}, slopes of
$I_d$ versus the gate ($V_g$) and drain ($V_d$) voltages (sections
\ref{subsection:poly} - \ref{subsection:IdvsVd}), role of anisotropic
effective mass (section \ref{subsection:effecmass}), role of gate
tunneling current as a function of gate oxide thickness and gate length
in determining the OFF current (section \ref{subsection:gateleakage}).
It is emphasized that the calculations presented include a
self-consistent treatment of
two dimensional gate oxide tunneling. Prior treatments of gate oxide
tunneling in semi-classical 2D simulators incorporated 1D models.

\section{Formalism}
\label{section:Formalism}

\subsection{The governing equations}

We consider $N_b$ independent valleys for the electrons within the
effective mass approximation. The Hamiltonian of valley $b$ is
\begin{eqnarray}
H_b(\vec{r}) = -\frac{\hbar^2}{2} [\frac{d}{dx} \left(\frac{1}{m^b_x}
\frac{d}{dx}\right) + \frac{d}{dy} \left(\frac{1}{m^b_y}\frac{d}{dy}
\right) + \frac{d}{dz} \left( \frac{1}{m^b_z}\frac{d}{dz} \right) ]
+ V(\vec{r}) \mbox{,}
\end{eqnarray}
where $(m^b_x,m^b_y,m^b_z)$ are the components of the effective
mass in valley $b$. The equation of motion for the retarded ($G^r$)
and less-than ($G^<$) Green's functions
are~\cite{Datta_book,Mahan_book,Mahan87}
\begin{eqnarray}
[E - H_{b_1} (\vec{r}_1)] G^r_{b_1,b_2} (\vec{r}_1,\vec{r}_2,E)
- \int d\vec{r} \; \Sigma^r_{b_1,b^\prime} (\vec{r}_1,\vec{r},E)
            G^r_{b^\prime,b_2} (\vec{r},\vec{r}_2,E) =
            \delta_{b_1,b_2}  \delta(\vec{r}_1-\vec{r}_2) \label{eq:Gr1}
\end{eqnarray}
and
\begin{eqnarray}
& & [E - H_{b_1} (\vec{r}_1)] G^<_{b_1,b_2} (\vec{r}_1,\vec{r}_2,E)
- \int d\vec{r} \; \Sigma^r_{b_1,b^\prime} (\vec{r}_1,\vec{r},E)
      G^<_{b^\prime,b_2} (\vec{r},\vec{r}_2,E) =  \nonumber  \\
& &
\hspace{3in}
\int d\vec{r} \; \Sigma^<_{b_1,b^\prime} (\vec{r}_1,\vec{r},E)
          G^a_{b^\prime,b_2} (\vec{r},\vec{r}_2,E) \label{eq:G<1}
\mbox{,}
\end{eqnarray}
where $G^a$ is the advanced Green's function. In the above equations,
the coordinate spans only the device (see Fig. \ref{fig:grid}).
The influence of the semi-infinite regions of the source (S), drain (D)
and polysilicon gate (P), and scattering mechanisms (electron-phonon)
are included via the self-energy terms $\Sigma^r_{b_1,b^\prime}$ and
$\Sigma^<_{b_1,b^\prime}$. We assume that charge is injected
independently from the contact into each valley. Then,
$\Sigma^\alpha_{b_1,b_2,C} = \Sigma^\alpha_{b_1,C} \;\delta_{b_1,b_2}$,
where $C$ represents the self-energy due to contacts.
Finally, the hole bands are treated using the drift-diffusion model,
which is expected to be a good approximation for n-MOSFETs.

The electrostatic potential varies in the $(x,y)$ plane, and the system
is translationally invariant along the z-axis.  So, all quantities
$A(\vec{r}_1,\vec{r}_2,E)$ depend only on the difference coordinate
$z_1-z_2$. Using the relation
\begin{eqnarray}
A(\vec{r}_1,\vec{r}_2,E) = \int \frac{d k_z}{2\pi} e^{ik_z (z_1 - z_2)}
                                    A(x_1,y_1,x_2,y_2,k_z,E) \mbox{ ,}
\end{eqnarray}
the equations of motion for $G^r$ and $G^<$ simplify to
\begin{eqnarray}
[E - \frac{\hbar^2 k_z^2}{2m_z} - H_b (\vec{r}_1)]
                  G^r_{b} (\vec{r}_1,\vec{r}_2,k_z,E)
- \int d\vec{r} \; \Sigma^r_{b} (\vec{r}_1,\vec{r},k_z,E)
            G^r_{b} (\vec{r},\vec{r}_2,k_z,E) =
                     \delta(\vec{r}_1-\vec{r}_2) \label{eq:Gr2}
\end{eqnarray}
and
\begin{eqnarray}
[E - \frac{\hbar^2 k_z^2}{2m_z} - H_b (\vec{r}_1)]
                  G^r_{b} (\vec{r}_1,\vec{r}_2,k_z,E)
- \int d\vec{r} \; \Sigma^r_{b} (\vec{r}_1,\vec{r},k_z,E)
            G^<_{b} (\vec{r},\vec{r}_2,k_z,E) =
\;\;\;\;\;\;\;\;\; && \nonumber \\
\int d\vec{r} \; \Sigma^<_{b} (\vec{r}_1,\vec{r},k_z,E)
        G^a_{b} (\vec{r},\vec{r}_2,k_z,E)\mbox{,}  &&
\;\;\;\;\;\;\;
\label{eq:G<2}
\end{eqnarray}
where $Z_b = Z_{b,b}$, and for the remainder of the paper,
$\vec{r} \rightarrow (x,y)$.

The density of states [$N(\vec{r},k_z,E)$] and charge density
[$\rho(\vec{r},k_z,E)$] are the sum of the contributions from the
individual valleys:
\begin{eqnarray}
N(\vec{r},k_z,E) &=& \sum_b N_b(\vec{r},k_z,E) = - \frac{1}{\pi}
      \mbox{Im}[ G_{b}^r(\vec{r},\vec{r},k_z,E) ] \label{eq:DOS}  \\
\rho(\vec{r},k_z,E) &=& \sum_b \rho_b(\vec{r},k_z,E) =
           -i G_{b}^<(\vec{r},\vec{r},k_z,E) \label{eq:density}
\mbox{ .}
\end{eqnarray}

\subsection{$G^r$ and $G^<$: Discretized matrix equations}
\label{subsection:algorithm}

Self-consistent solution of the Green's function and Poisson's
equations requires repeated computation of the non-equilibrium charge
density. This computation is often the most time consuming
part in modeling the electronic properties of devices.

The common procedure to evaluate the electron density uses the
expression
\begin{eqnarray}
\rho_b(\vec{r},k_z,E) &=& -i G^<_{b}(\vec{r},\vec{r},k_z,E) \nonumber \\
      &=& -i \int d\vec{r}_1 d\vec{r}_2 G^r_{b}(\vec{r},\vec{r}_1,k_z,E)
                   \Sigma^<_{b}(\vec{r}_1,\vec{r}_2,k_z,E) 
            G^a_{b}(\vec{r}_2,\vec{r},k_z,E) \mbox{,} \label{eq:grsigga}
\end{eqnarray}
where  $G^r(\vec{r}_1,\vec{r}_2,k_z,E)$ must be computed between all
$N_x N_y$ grid
points and those grid points involving a non zero $\Sigma^\alpha$. The
operation count required to solve for all elements of $G^r$ scales as
$(N_x N_y)^3$, and so the use of Eq. (\ref{eq:grsigga}) is expensive.
We have developed a new recursive algorithm to compute the electron 
density in systems when the discretized version of the LHS of Eqs.
(\ref{eq:Gr2}) and (\ref{eq:G<2}) is block tridiagonal. This algorithm
requires only the evaluation of the diagonal blocks of $G^r$. The
operation count of this algorithm scales as $N_x^3 N_y$ (or $N_x N_y^3$)
when the diagonal blocks correspond to lattice points in the $x$ (or
$y$) direction. We summarize the recursive algorithm to calculate $G^r$
(section \ref{subsection:algorithm_Gr}) as it sets the stage for the
new algorithm to compute $G^<$ (section \ref{subsection:algorithm_G<}).
We stress that Poisson's equation only requires the diagonal elements
of $G^<$ (Eq. (\ref{eq:grsigga})). The algorithm we develop in section
\ref{subsection:algorithm_Gr} however computes the diagonal blocks of
$G^<$. While this is much better than using Eq. (\ref{eq:grsigga})
directly as discussed above, new algorithms to solve for only the
diagonal elements with operation counts smaller than $N_x^3 N_y$ (or
$N_x N_y^3$) are very desirable.

In matrix form, Eqs. (\ref{eq:Gr2}) and (\ref{eq:G<2}) are written as
\begin{eqnarray}
A^\prime G^r &=& \lambda \mbox{ and } \label{eq:discreteGr1} \\
\mbox{ and } \;\;\;\;\;\;\;\
A^\prime G^< &=& \Sigma^< G^a \mbox{ . } \label{eq:discreteG<1}
\end{eqnarray}
 The self-energies due to the S, D and P are non zero only along the
lines $y=y_S=y_1$, $y=y_D=y_{N_y}$ and $x=x_P$ respectively (see
Fig. \ref{fig:grid}). The $A^\prime$ matrix has a dimension of
$N_x N_y$ and is ordered such that all grid points located at a
particular y-coordinate correspond to its diagonal blocks. The notation
adopted is that $A^\prime_{j_1,j_2}(i,i^\prime)$ refers to the
off-diagonal entry corresponding to grid points $(x_i,y_{j_1})$ and
$(x_i^\prime,y_{j_2})$. The non zero elements of the diagonal blocks of
the $A^\prime$ matrix are given by
\begin{eqnarray}
A^\prime_{j,j}(i,i) \!\! &=& \!\! E^\prime - V_{i,j}  - T_{j,j} (i+1,i)
                - T_{j,j} (i-1,i) - T_{j+1,j} (i,i) - T_{j-1,j} (i,i)
                                                        \nonumber\\
           \!\!  & & \!\!
- \Sigma^r_S(x_i,x_i) \delta_{j,1} - \Sigma^r_D(x_i,x_i) \delta_{j,N_y}
- \Sigma^r_P(y_j,y_j) \delta_{i,1} -\Sigma^r (x_i,y_j;x_i,y_j)  \\
A^\prime_{j,j}(i\pm1,i) \!\! &=& \!\! T_{j,j} (i\pm1,i)
- \Sigma^r_S(x_{i\pm1},x_i) \delta_{j,1} - \Sigma^r_D(x_{i\pm1},x_i)
  \delta_{j,N_y} \nonumber \\
             & & \;\;\;\;\;\;\;\;\;\;\;\;\;\;\;\;\;\;\;\;\;\;\;\;\;\;\;
\;\;\;\;\;\;\;\;\;\;\;\;\;\;\;\;\;\;\;\;\;\;\;\;\;\;\;\;\;\;\;\;\;\;\;\;
-\Sigma^r (x_{i\pm1},y_j;x_i,y_j) \\
A^\prime_{j,j}(i,i^\prime) \!\! &=& \!\! - \Sigma^r_S(x_i,x_{i^\prime})
            \delta_{j,1} - \Sigma^r_D(x_i,x_{i^\prime}) \delta_{j,N_y}
\mbox{, for $i^\prime \neq i,\;i\pm1$ ,}
\end{eqnarray}
where  $E^\prime = E - \hbar^2 k_z^2 / 2m_z$ and $V_{i,j}=V(x_i,y_j)$.
The off-diagonal blocks are
\begin{eqnarray}
A^\prime_{j\pm1,j}(i,i) &=& T_{j\pm1,j} (i,i)
               - \Sigma^r_P(y_j,y_{j\pm1}) \delta_{i,1} \nonumber \\
A^\prime_{j,j^\prime} (i,i^\prime) &=& 0 \mbox{, for $j^\prime \neq j,\;
                                                j\pm1$.}
\end{eqnarray}
The non zero elements of the $T$ matrix are defined by
\begin{eqnarray}
T_{j,j} (i\pm1,i) & = & \frac{\hbar^2}{2m^{\pm x}}
                  \frac{2}{x_{i+1}-x_{i-1}} \frac{1}{|x_{i\pm1}-x_i|}
                        \label{eq:T1}  \\
T_{j\pm1,j} (i,i) & = & \frac{\hbar^2}{2m^{\pm y}}
                  \frac{2}{y_{j+1}-y_{j-1}} \frac{1}{|y_{j\pm1}-y_j|}
                                                        \mbox{,}
                        \label{eq:T2}
\end{eqnarray}
where $m^{\pm x}=\frac{2}{m_{i\pm 1,j} + m_{i,j}}$ and
$m^{\pm y}=\frac{2}{m_{i,j\pm 1} + m_{i,j}}$.
Non zero elements of $\Sigma^r_P(y_j,y_j^\prime)$, where $j^\prime \neq
j$ are neglected to ensure that $A^\prime$ is block tridiagonal (the
algorithm to calculate $G^r$ and $G^<$ relies on the block tridiagonal
form of $A^\prime$). The $\lambda$ appearing in Eq.
(\ref{eq:discreteGr1}) corresponds to the delta function in Eq.
(\ref{eq:Gr2}). $\lambda$ is a diagonal matrix whose elements are
given by
\begin{eqnarray}
\lambda_{i,j;i,j} = \frac{4}{(x_{i+1}-x_{i-1}) (y_{i+1}-y_{i-1})}
\mbox{ .}
\end{eqnarray}

\subsection{Recursive algorithm to calculate $G^r$}
\label{subsection:algorithm_Gr}

Pre-multiplying Eq. (\ref{eq:discreteGr1}) by $\lambda^{-1}$,
\begin{eqnarray}
A \; G^r = I \mbox{ ,} \label{eq:discreteGr2}
\end{eqnarray}
where  matrix $A$ is a symmetric matrix for both uniform and non
uniform rectangular grids (Note that $A^\prime$ is symmetric only for
a uniform grid). The recursive algorithm to calculate the diagonal
blocks of the full Green's function is discussed now, using Dyson's
equation for $G^r$, and the left-connected Green's function as in
references:~\cite{Lake97,NEMO}

\noindent
(i) \underline{Dyson's equation for $G^r$}:
The solution to
\begin{eqnarray}
\tiny{
\left(
\begin{array}{cc}
A_{Z,Z}    & A_{Z,Z^\prime}  \\
A_{Z^\prime,Z} & A_{Z^\prime,Z^\prime}
\end{array}
\right)
\left(
\begin{array}{cc}
G^r_{Z,Z} & G^r_{Z,Z^\prime} \\
G^r_{Z^\prime,Z} & G^r_{Z^\prime,Z^\prime}
\end{array}
\right)
=
\left(
\begin{array}{cc}
I  & O         \\
O  & I
\end{array}
\right)
} \mbox{,} \label{eq:Gr}
\end{eqnarray}
is
\begin{eqnarray}
G^r &=& G^{r0} + G^{r0} U G^r  \label{eq:dysonGr2} \\
    &=& G^{r0} + G^r U G^{r0} \mbox{  ,} \label{eq:dysonGr3}
\end{eqnarray}
where,
\begin{eqnarray}
G^{r}=
\tiny{ \left(
\begin{array}{cc}
G^{r}_{Z,Z}        & G^{r}_{Z,Z^\prime}         \\
G^{r}_{Z^\prime,Z} & G^{r}_{Z^\prime,Z^\prime}
\end{array}
\right)}
\mbox{, }
G^{r0}=
{\tiny{ \left(
\begin{array}{cc}
G^{r0}_{Z,Z}  & O         \\
O  & G^{r0}_{Z^\prime,Z^\prime}
\end{array}
\right)}
=
\tiny{ \left(
\begin{array}{cc}
A_{Z,Z}^{-1}  & O         \\
O  & A_{Z^\prime,Z^\prime}^{-1}
\end{array}
\right) }}
\mbox{ and }
U=
\tiny{ \left(
\begin{array}{cc}
O    & - A_{Z,Z^\prime}  \\
- A_{Z^\prime,Z} & O
\end{array}
\right) }
\mbox{ .} \label{eq:GrandU}
\end{eqnarray}

The advanced Green's function ($G^a$) is by definition related to $G^r$
by
\begin{eqnarray}
G^a = {G^{r}}^\dagger &=& G^{a0} + G^{a0} U^\dagger G^a
            \label{eq:Ga1} \\
                      &=& G^{a0} + G^{a} U^\dagger G^{a0}
 \mbox{  .} \label{eq:Ga2}
\end{eqnarray}
Eq. (\ref{eq:dysonGr2}) is called Dyson's
equation.~\cite{Datta_book,Mahan_book}

\noindent
(ii) \underline{Left-connected retarded Green's function}:
The left-connected (superscript $L$) retarded (superscript $r$) Green's
function $g^{rLq}$ is defined by the first $q$ blocks of
Eq. (\ref{eq:discreteGr2}) (includes the open boundaries attached
to the lattice points via the self-energy) by
\begin{eqnarray}
A_{1:q,1:q} \; g^{rLq} = I_{q,q} \mbox{, where, }
I_q=I_{1:q,1:q} \mbox{ .}
\end{eqnarray}
$g^{rLq+1}$ is defined in a manner identical to $g^{rLq}$ except that
the left-connected system is comprised of the first $q+1$ blocks of
Eq. (\ref{eq:discreteGr2}). In terms of Eq. (\ref{eq:Gr}), the
equation governing $g^{rLq+1}$ follows by setting $Z=1:q$ and
$Z^\prime=q+1$.
Using Dyson's equation [Eq. (\ref{eq:dysonGr2})], we obtain
\begin{eqnarray}
g^{rLq+1}_{q+1,q+1}= \left(A_{q+1,q+1} +
             A_{q+1,q} g^{rLq}_{q,q}
          A_{q,q+1} \right)^{-1} \mbox{ .} \label{eq:dysongrL2}
\end{eqnarray}
Note that the last element $g^{rLN}_{N,N}$ is equal to the fully
connected Green's function $G^r_{N,N}$, which is the solution to
Eq. \ref{eq:discreteGr2}.

\noindent
(iii) \underline{Full Green's function in terms of the left-connected
Green's function}:
Consider Eq. (\ref{eq:Gr}) such that $A_{Z,Z}=A_{1:q,1:q}$,
$A_{Z^\prime,Z^\prime}=A_{q+1:N,q+1:N}$ and
$A_{Z,Z^\prime}=A_{1:q,q+1:N}$. Noting that the only nonzero element
of $A_{1:q,q+1:N}$ is $A_{q,q+1}$ and using Eq. (\ref{eq:dysonGr2}),
we obtain
\begin{eqnarray}
G^r_{q,q} &=& g^{rLq}_{q,q} + g^{rLq}_{q,q}
             \left(A_{q,q+1} G^r_{q+1,q+1} A_{q+1,q} \right)
                    g^{rLq}_{q,q} \label{eq:discreteGr3} \\
          &=& g^{rLq}_{q,q} + g^{rLq}_{q,q} A_{q,q+1} G^r_{q+1,q}
                    \label{eq:discreteGr4}
\mbox{ .}
\end{eqnarray}
Both $G^r_{q,q}$ and $G^r_{q+1,q}$ are used in the algorithm for
electron density, and so storing both sets of matrices will be
useful.

In view of the above equations, the algorithm to compute the diagonal
blocks $G^r_{q,q}$ is given by the following steps:
\begin{itemize}
\item $g^{rL1}_{11} = A_1^{-1}$.
\item For $q=1,2, ...,N-1$, compute Eq. (\ref{eq:dysongrL2}).
\item For $q=N-1, N-2, ..., 1$, compute Eq. (\ref{eq:discreteGr4}).
Store $G^r_{q+1,q}$ if memory permits for use in the algorithm
for electron density.
\end{itemize}

\subsection{Recursive algorithm to calculate density ($G^<$)}
\label{subsection:algorithm_G<}

The discretized form of Eq. (\ref{eq:G<2}) is
\begin{eqnarray}
A^\prime G^< = \Sigma^< G^a \mbox{,} \label{eq:discreteG<1p}
\end{eqnarray}
where the dimension of the matrices involved are $N=N_x N_y$.
Pre-multiplying by $\lambda^{-1}$,
\begin{eqnarray}
A G^< = \Sigma^< G^a  \mbox{ ,} \label{eq:discreteG<2}
\end{eqnarray}
where $\Sigma^<$ in Eq. (\ref{eq:discreteG<2}) is equal to
$\lambda^{-1}$ times the $\Sigma^<$ that appears in Eqs.
(\ref{eq:G<1}) and (\ref{eq:discreteG<1p}).

Following subsection \ref{subsection:algorithm_Gr}, the algorithm to
calculate the electron density (diagonal elements of $G^<$) is
discussed in terms of a Dyson's equation for $G^<$ and the
left-connected $g^{<L}$:

\noindent
(i) \underline{Dyson's equation for $G^<$}:
The solution to
\begin{eqnarray}
\tiny{
\left(
\begin{array}{cc}
A_{Z,Z}    & A_{Z,Z^\prime}  \\
A_{Z^\prime,Z} & A_{Z^\prime,Z^\prime}
\end{array}
\right)
\left(
\begin{array}{cc}
G^<_{Z,Z}         & G^<_{Z,Z^\prime} \\
G^<_{Z^\prime,Z}  & G^<_{Z^\prime,Z^\prime}
\end{array}
\right)
=
\left(
\begin{array}{cc}
\Sigma^<_{Z,Z}         & \Sigma^<_{Z,Z^\prime}  \\
\Sigma^<_{Z^\prime,Z}  & \Sigma^<_{Z^\prime,Z^\prime}
\end{array}
\right)
\left(
\begin{array}{cc}
G^a_{Z,Z}         & G^a_{Z,Z^\prime} \\
G^a_{Z^\prime,Z}  & G^a_{Z^\prime,Z^\prime}
\end{array}
\right)
} \mbox{ } \label{eq:discreteG<3}
\end{eqnarray}
can be written as
\begin{eqnarray}
G^< &=& G^{r0} U G^< + G^{r0} \Sigma^< G^a \mbox{,} \label{eq:dysonG<1}
\end{eqnarray}
where $G^{r0}$ and $U$ have been defined in Eqs. (\ref{eq:GrandU}),
and $G^<$ and $G^a$ are readily identifiable from
Eq. (\ref{eq:discreteG<3}). Using $G^a=G^{a0} + G^{a0} U^\dagger G^{a}$,
Eq. (\ref{eq:dysonG<1}) can be written as
\begin{eqnarray}
G^< &=& G^{<0} + G^{<0} U^\dagger G^a + G^{r0} U G^<
                                        \label{eq:dysonG<3}\\
    &=& G^{<0} + G^r U G^{<0} + G^< U^\dagger G^{a0}
                                        \label{eq:dysonG<4}
                                                     \mbox{, } \\
 \mbox{where  }
G^{<0} &=& G^{r0} \Sigma^< G^{a0}  \mbox{ .}  \label{eq:G<0}
\end{eqnarray}

\noindent
(ii) \underline{Left-connected $g^<$}:
$g^{<Lq}$ is the counter part of $g^{rLq}$, and is
defined by the first $q$ blocks of Eq. (\ref{eq:discreteG<2}):
\begin{eqnarray}
A_{1:q,1:q} \; g^{<Lq} =
\Sigma^<_{1:q,1:q} \; g^{aLq}_{1:q,1:q} \mbox{ .}
\end{eqnarray}
$g^{<Lq+1}$ is defined in a manner identical to $g^{<Lq}$ except that
the left-connected system is comprised of the first $q+1$ blocks of
Eq. (\ref{eq:discreteG<2}). In terms of Eq. (\ref{eq:discreteG<3}), the
equation governing $g^{<Lq+1}$ follows by setting $Z=1:q$ and
$Z^\prime=q+1$. Using the Dyson's equations for $G^r$ and $G^<$,
$g^{<Lq+1}_{q+1,q+1}$ can be recursively obtained (derivation
is presented in Appendix A) as
\begin{eqnarray}
g^{<Lq+1}_{q+1,q+1}
=
g^{rLq+1}_{q+1,q+1} \left[\Sigma^{<}_{q+1,q+1}
+ \sigma^<_{q+1} \right] g^{aLq+1}_{q+1,q+1}
+ g^{rLq+1}_{q+1,q+1} \; \Sigma^{<}_{q+1,q} \;
g^{aLq+1}_{q,q+1}
+ g^{rLq+1}_{q+1,q} \; \Sigma^{<}_{q,q+1} \;
g^{aLq+1}_{q+1,q+1} \mbox{,} \label{eq:g<L2}
\end{eqnarray}
which can be written in a more intuitive form as
\begin{eqnarray}
g^{<Lq+1}_{q+1,q+1}
=
g^{rLq+1}_{q+1,q+1} \left[
\Sigma^{<}_{q+1,q+1} + \sigma^<_{q+1}
+ \Sigma^{<}_{q+1,q} \; g^{aLq}_{q,q} \;
A_{q,q+1}^\dagger
+ A_{q+1,q} \; g^{rLq}_{q,q}\; \Sigma^{<}_{q,q+1}
\right] g^{aLq+1}_{q+1,q+1} \mbox{,} \label{eq:g<L}
\end{eqnarray}
where
$\sigma^<_{q+1} = A_{q+1,q} g^{<Lq}_{q,q} A_{q,q+1}^\dagger$.
Eq. (\ref{eq:g<L}) has the physical meaning that
$g^{<Lq+1}_{q+1,q+1}$ has contributions due to four
injection functions: (i) an effective self-energy due to the
left-connected structure that ends at $q$, which is represented by
$\sigma^<_{q+1}$,  (ii) the diagonal self-energy component at grid
point $q+1$ that enters Eq. (\ref{eq:discreteG<2}), and (iii) the two
off-diagonal self-energy components involving grid points $q$ and $q+1$.

\noindent
(iii) \underline{Full less-than Green's function in terms of
left-connected Green's function}:
Consider Eq. (\ref{eq:discreteG<3}) such that $A_Z=A_{1:q,1:q}$,
$A_Z^\prime=A_{q+1:N,q+1:N}$ and $A_{Z,Z^\prime}=A_{1:q,q+1:N}$.
Noting that the only nonzero element of $A_{1:q,q+1:N}$ is
$A_{q,q+1}$ and using Eq. (\ref{eq:dysonG<3}), we obtain
\begin{eqnarray}
G^<_{q,q} = g^{<Lq}_{q,q}
                 + g^{<Lq}_{q,q} A_{q,q+1}^\dagger G^a_{q+1,q}
                 + g^{<0}_{q,q+1} A_{q+1,q}^\dagger G^a_{q,q}
                 + g^{rLq}_{q,q} A_{q,q+1} G^<_{q+1,q}
                         \mbox{ .} \label{eq:discreteG<4}
\end{eqnarray}
Using Eq. (\ref{eq:dysonG<4}), $G^<_{q+1,q}$ can be written in terms
of $G^<_{q+1,q+1}$ and other known Green's functions as
\begin{eqnarray}
G^<_{q+1,q} = g^{<0}_{q+1,q}
                 + G^r_{q+1,q} A_{q,q+1} g^{<0}_{q+1,q}
                 + G^r_{q+1,q+1} A_{q+1,q} g^{<Lq}_{q,q}
           + G^<_{q+1,q+1} A_{q,q+1}^\dagger g^{aLq}_{q,q}
                    \mbox{ .}   \label{eq:discreteG<5}
\end{eqnarray}
Substituting Eq. (\ref{eq:discreteG<5}) in Eq. (\ref{eq:discreteG<4})
and using Eqs. (\ref{eq:dysonGr2}) and (\ref{eq:dysonGr3}), we obtain
\begin{eqnarray}
G^<_{q,q}&=&g^{<Lq}_{q,q} +
            g^{rLq}_{q,q} \left(A_{q,q+1} G^<_{q+1,q+1}
             A_{q+1,q}^\dagger \right) g^{aLq}_{q,q} +
            \left[g^{<Lq}_{q,q} A_{q,q+1}^\dagger G^a_{q+1,q} +
             G^r_{q,q+1} A_{q+1,q} g^{<Lq}_{q,q} \right]
             \nonumber \\
         & & \;\;\;\;\;\;\; \;\;\;\;\;\;\; \;\;\;\;\;\;\;
            + \left[g^{<0}_{q,q+1} A_{q+1,q}^\dagger G^a_{q,q} +
             G^r_{q,q} A_{q,q+1} g^{<0}_{q+1,q} \right]
                         \mbox{,} \label{eq:discreteG<6}
\end{eqnarray}
where
\begin{eqnarray}
g^{<0}_{q,q+1} &=& g^{r0}_{q,q} \Sigma^<_{q,q+1} g^{a0}_{q+1,q+1} \\
g^{<0}_{q+1,q} &=& g^{r0}_{q+1,q+1} \Sigma^<_{q+1,q} g^{a0}_{q,q}
                       \mbox{ .}                   \label{eq:g4temp}
\end{eqnarray}
The terms inside the square brackets of Eq. (\ref{eq:discreteG<6})
are Hermitian conjugates of each other.

In view of the above equations, the algorithm to compute the diagonal
blocks of $G^<$ is given by the following steps:
\begin{itemize}
\item $g^{<L1}_{11} = g^{r0}_{11} \Sigma_L^< g^{a0}_{11}$.
\item For $q=1,2, ...,N-1$, compute Eq. (\ref{eq:g<L}).
\item For $q=N-1, N-2, ..., 1$, compute Eqs. (\ref{eq:discreteG<6})
- (\ref{eq:g4temp}).
\end{itemize}

The current density flowing between two neighboring blocks $q$ and
$q+1$ is given by
\begin{eqnarray}
J(q \rightarrow q+1,k_z,E) &=& \sum_b J_b(q \rightarrow q+1,k_z,E)
                                                       \nonumber \\
                         &=&
\frac{2e}{\hbar} \sum_b \mbox{Tr}
\left[
T_{q,q+1} G^<_{b;q,q+1} (k_z,E) - T_{q+1,q} G^<_{b;q+1,q} (k_z,E)
\right ] \mbox{,}
\end{eqnarray}
where $T$ has been defined in Eqs. (\ref{eq:T1}) and (\ref{eq:T2}).
The current that has leaked into the gate between any two blocks $p$
and $q$ is
\begin{eqnarray}
J_{gate}^{qp}  &=& \sum_b J_b(p \rightarrow p+1,k_z,E) -
\sum_b J_b(q \rightarrow q+1,k_z,E) \mbox{ ,}
\end{eqnarray}
and the total gate leakage current obtained by choosing $p$ and $q$
near the source and drain ends of the device.

\subsection{Expressions for Contact Self-energies
($\Sigma^r_S$, $\Sigma^r_D$ and $\Sigma^r_P$)}
\label{subsection:contact_selfenergies}

Potential and doping profiles in the semi-infinite regions to
the (a) left of 'S' and right of 'D' are equal to the value at q = 1
and $N_y$ respectively (Fig. \ref{fig:grid}). That is, they do not vary as
a
function of the y-coordinate, and (b) top of the 'P' is equal to the
value of the top most grid line of 'P' (Fig. \ref{fig:grid}). That is, they
are not
a function of the x-coordinate. The retarded surface Green's functions
of these semi-infinite regions are calculated from
Eq. (\ref{eq:discreteGr2}), when the matrices involved are
semi-infinite. All diagonal sub-matrices of the $A$ matrix are equal to
$A_{1,1}$, $A_{N_y,N_y}$ and $A_P$, and all first upper off-diagonal
matrices of the $A$ matrix are equal to $A_{1,2}$, $A_{N_y-1,N_y}$ and
$A_{P-1,P}$, in the source, drain and polysilicon regions respectively.
We spell out the entire matrix for the source semi-infinite regions
below:
\begin{eqnarray}
\tiny{ \left(
\begin{array}{ccccccc}
\bullet & \bullet & 0       & 0       & 0       & 0       & 0 \\
\bullet & \bullet & \bullet & 0       & 0       & 0       & 0 \\
0       & \bullet & \bullet & \bullet & 0       & 0       & 0 \\
0       & 0       & A_{2,1} & A_{1,1} & A_{1,2} & 0       & 0 \\
0       & 0       & 0       & A_{2,1} & A_{1,1} & A_{1,2} & 0 \\
0       & 0       & 0       & 0       & A_{2,1} & A_{1,1} & A_{1,2}\\
0       & 0       & 0       & 0       & 0       & A_{2,1} & A_{1,1}
\end{array}
\right)}
\tiny{ \left(
\begin{array}{ccccccc}
\bullet & \bullet & \bullet   & \bullet   & \bullet   & \bullet   & \bullet
\\^M\bullet & \bullet & \bullet   & \bullet   & \bullet   & \bullet   &
\bullet \\^M\bullet & \bullet & \bullet   & \bullet   & \bullet   & \bullet
& \bullet \\^M\bullet & \bullet & \bullet   & g_{-3,-3} & g_{-3,-2} &
g_{-3,-1} & g_{-3,0} \\
\bullet & \bullet & \bullet   & g_{-2,-3} & g_{-2,-2} & g_{-2,-1} & g_{-2,0}
\\
\bullet & \bullet & \bullet   & g_{-1,-3} & g_{-1,-2} & g_{-1,-1} &
g_{-1,0}\\^M\bullet & \bullet & \bullet   & g_{0,-3}  & g_{0,-2}  & g_{0,-1}
& g_{0,0}
\end{array}
\right)}
=
\tiny{ \left(
\begin{array}{ccccccc}
\bullet & 0 & 0 & 0 & 0 & 0 & 0 \\
0 & \bullet & 0 & 0 & 0 & 0 & 0 \\
0 & 0 & \bullet & 0 & 0 & 0 & 0 \\
0 & 0 & 0 & I & 0 & 0 & 0 \\
0 & 0 & 0 & 0 & I & 0 & 0 \\
0 & 0 & 0 & 0 & 0 & I & 0 \\
0 & 0 & 0 & 0 & 0 & 0 & I
\end{array}
\right)} \mbox{.}
\end{eqnarray}
The surface Green's function of these regions can be obtained by using
methods in matrix algebra that transform the two dimensional wire 
representing the semi-infinite contacts with $N_x$ grid points to 
$N_x$ one dimensional wires.

The self-energies due to the contacts are:
\begin{eqnarray}
\Sigma^r_S(k_z,E) &=& A_{1,0} g_{0,0} (k_z,E) A_{0,1} \\
\Sigma^r_D(k_z,E) &=& A_{N_y,N_y+1} g_{N_y+1,N_y+1}(k_z,E)
                                              A_{N_y+1,N_y} \\
\Sigma^r_P(k_z,E) &=& A_{P} g_{P} (k_z,E) A_{P} \\
\Sigma^<_S(k_z,E) &=& - 2 i A_{1,0} Im \left[g_{0,0} (k_z,E) \right]
                                                  A_{0,1} f_S(E) \\
\Sigma^<_D(k_z,E) &=& - 2 i A_{N_y,N_y+1}
        Im \left[g_{N_y+1,N_y+1} (k_z,E) \right] A_{N_y+1,N_y} f_D(E) \\
\Sigma^<_P(k_z,E) &=& - 2 i A_{P} Im \left[g_{P} (k_z,E) \right] A_{P}
                                                               f_P(E)
\mbox{,}
\end{eqnarray}
where  $f_i(E)$ is the Fermi factor in contact $i \in S,\; D,\; P$.

When $\Sigma^{\alpha}_{b} (\vec{r}_1,\vec{r}_2,k_z,E)$ depends only on
$E_{xy} = E - \frac{\hbar^2 k_z^2}{2 m_z}$, then Eqs. (\ref{eq:Gr2})
and (\ref{eq:G<2}) simplify to
\begin{eqnarray}
[E_{xy} - H_b (\vec{r}_1)]
                    G^r_{b} (\vec{r}_1,\vec{r}_2,E_{xy})
  - \int d\vec{r} \; \Sigma^r_{b} (\vec{r}_1,\vec{r},E_{xy})
              G^r_{b} (\vec{r},\vec{r}_2,E_{xy}) =
                       \delta(\vec{r}_1-\vec{r}_2) \\
\end{eqnarray}
and
\begin{eqnarray}
&& [E_{xy} - H_{b} (\vec{r}_1)]
                   G^<(\vec{r}_1,\vec{r}_2,E_{xy})
  - \int d\vec{r} \; \Sigma^r_{b} (\vec{r}_1,\vec{r},E_{xy})
               G^<_{b} (\vec{r},\vec{r}_2,k_z,E) =  \nonumber \\
&& \;\;\;\;\;\;\;\;\;\;\;\;\;\;\;\;\;\;\;\;\;\;\;\;
  \;\;\;\;\;\;\;\;\;\;\;\;\;\;\;\;\;\;\;\;\;\;\;\;\;\;\;
  \int d\vec{r} \; \Sigma^<_{b} (\vec{r}_1,\vec{r},E_{xy})
             G^a_{b} (\vec{r},\vec{r}_2,E_{xy})
\mbox{.}
\end{eqnarray}
While solving the equations, to keep the problem two dimensional,
$m_z$ has to be independent of (x,y). So, we assume $m_z$(SiO$_2$)
= $m_z$(Si).

\section{Results and Discussion}

The steady state characteristics of MOSFETS that are of practical
interest are the drive current, OFF current, slope of drain current
versus drain voltage, and threshold voltage. In this section, we show
that quantum mechanical simulations yield significantly different
results from drift-diffusion based methods. These differences arise
because of the following quantum mechanical features:

\noindent
(i) polysilicon gate depletion in a manner opposite to the classical
    case,

\noindent
(ii) dependence of the resonant levels in the channel on the gate
     voltage,

\noindent
(iii) tunneling of charge across the gate oxide and from source to
      drain,

\noindent
(iv) quasi-ballistic flow of electrons.




The MIT well-tempered 25 nm device structure~\cite{MIT25inputdeck} is
chosen for the purpose of discussion (MIT 25 nm device 
structure~\cite{MIT25inputdeck} is hereafter referred to as MIT25). The
method and computer code developed can however handle a wide variety 
of two dimensional structures with many terminals.
We first compare the potential profiles from a constant mobility
drift-diffusion solution and our quantum calculations at equilibrium.
The motivation for this comparison results from the observation
that the classical and quantum potential profiles should be in
reasonable agreement, if the doping density is significantly higher
than the electron and hole densities and the boundary conditions are
the same. The doping profile of MIT25 meets this requirement in the
channel region at small $V_g$, and we verify that the potential
profiles are in reasonable agreement at $y=0$ (see 'Q1 flat band' and
'DD flat band' of Fig. \ref{fig:PolyFlatBand}). The legend 'flat band'
refers to the potential at $x=-t_{ox}$ being fixed at the applied gate
potential.

An index of abbreviations used follows:

\noindent
Length Scales:
$t_{ox}$ - oxide thickness,
$L_P$ - polysilicon gate thickness in x-direction,
$L_B$ - boundary of substrate region in x-direction,
$L_y$ - Poisson's and NEGF equations are solved from $-L_y/2$ to
        $+L_y/2$,
$L_g$ - length of  polysilicon gate region in $y$-direction.

\noindent
Models:
Q1 - quantum mechanical calculations using an isotropic effective mass,
Q3 - quantum mechanical calculations using an anisotropic effective
      mass,
DD - drift diffusion,
Flat band - potential in the polysilicon gate region is held fixed
from $x=-(t_{ox}+L_P)$ to $x=-t_{ox}$ at the bulk value.
q-poly - potential in the gate polysilicon region is held fixed
at $x=-(t_{ox}+L_P)$ at the bulk value, and the potential is computed
quantum mechanically (self-consistently) for $x > -(t_{ox}+L_P)$.
c-poly - classical treatment of gate polysilicon region, as in DD.

\noindent
Current and voltage: $I_d$ - drain current, $I_g$ - gate current,
$V_d$ - drain voltage, $V_g$ - gate voltage.

The values of constants assumed to obtain the numerical results of this
section are as follows, unless otherwise noted:

Electron effective mass of silicon: 0.3283 (isotropic), 0.19 and
0.98 (anisotropic), Electron effective mass of SiO$_2$:
$m_x = m_y = 0.5$ and same as silicon in $m_z$ direction,
Hole effective mass of silicon is 0.49,
band gap of silicon (Si$O_2$): 1.12 eV (8.8 eV),
energy barrier between the  silicon and the oxide $\Delta E_C $=3.1 eV,
dielectric constant of Si (SiO$_2$) is  $\epsilon_{Si}$=11.9 (3.9) and
kT = 0.02585 eV.

\subsection{$I_d$ versus $V_g$ - Effect of polysilicon depletion region}
\label{subsection:poly}

The quantum mechanically calculated electron density near the SiO$_2$
barrier in the polysilicon region is smaller than the uniform
background doping density. This is because the electron wavefunction 
is small close to the barrier. As a result, the conduction
band in the polysilicon gate bends in a direction opposite to that
computed semi-classically (compare x and triangle in Fig.
\ref{fig:PolyFlatBand}).~\cite{Bowen,Spinelli00}
The band bending in the polysilicon gate plays
a significant role in determining the threshold voltage and
OFF current. To emphasize the importance of band bending, we plot
the drain current versus gate voltage calculated with the gate
polysilicon region treated as (i) 'flat band' and (ii) 'q-poly'.
We find that the computed current is larger in (ii) because quantum
mechanical depletion of electrons in the polysilicon gate region close
to the oxide causes lowering of the potential in the channel. The $I_d$
versus $V_g$ curve shifts by approximately the an amount equal to the
band bending in the polysilicon gate, in comparison to the flat band
case. This band bending, which is measured from $-(L_P+t_{ox})$ to
$-t_{ox}$ at equilibrium, is about 130 meV at the given doping density
(Fig.  \ref{fig:PolyFlatBand}). 
The influence of bandgap narrowing has been neglected here. It must be
mentioned that the bandgap narrowing effect will tend to make the 
quantum mechanical contribution to the polysilicon band bending just 
discussed smaller. Future work to determine the influence of bandgap 
narrowing is necessary.

Computationally, a 2D treatment of the polysilicon gate region is
expensive because of the additional grid points required. Note that
matrix inversion depends on the cube of the matrix dimension. We point
out that for highly doped polysilicon gate (in the absence of gate
tunneling), a shift in the $I_d (V_g)$ curve from (i) by the
equilibrium 1D built-in potential does a reasonable job of reproducing
the quantum mechanical result (see triangles in Fig.
\ref{fig:IdvsVg_1B_1BP}). This approximation becomes progressively
poorer with increase in gate voltage, as can be seen from the figure.
This is true espcecially for a smaller polysilicon doping density such
as 1E20.

\subsection{$I_d$ versus $V_g$ - Comparison to Medici}
\label{subsection:medici}

In the absence of gate tunneling and inelastic tunneling, the quantum
mechanical current is
\begin{eqnarray}
I_d = \frac{2e}{h}
     \int dE \; T_{SD} (E) \left[f_S (E) - f_D (E) \right] \mbox{,}
\end{eqnarray}
where $T_{SD}$ is the transmission probability from source to drain,
and $f_S$ and $f_D$ are the Fermi-Dirac factors in the source and drain
respectively. The total transmission (Fig. \ref{fig:DOSandTvsEbw}) is
step-like with integer values at the plateaus in-spite of the
complicated two dimensional electrostatics. In visual terms,
the energies at which the steps turn on are determined
by an effective 'subband dependent' source injection barrier, in
contrast to the source injection barrier in drift-diffusion
calculations.~\cite{Lundstrom97} This subband dependent source
injection barrier is simply
the maximum energy of the subband between source and drain due to
quantization in the direction perpendicular to the gate plane
(x-direction of Fig. 1). From a practical view point, the following
two issues are important in ballistic MOSFETs:
(a) typically, the total transmission assumes integer value at an
energy slightly above the maximum in 2D density of states as shown in
the inset of Fig. \ref{fig:DOSandTvsEbw}, and (b) the steps develop
over 50 meV (twice the room temperature thermal energy). So, the shape
of the steps is important in determining the value of current.
Assuming a sharp step in total
transmission with integer values in a calculation of current as in
reference \onlinecite{Natori94} is not quite accurate.

We compare the results from our quantum simulations with published
results from quantum-corrected Medici.~\cite{MIT25inputdeck}
To compare the quantum and classical results, an estimate of
the energy of the first subband minima ($E_{r1}$) from Fig.
\ref{fig:DOSandTvsEbw}, and the location of the classical barrier
height ($E_b(classical)$) (Fig. \ref{fig:EbERvsVg}) are
useful ($E_b(classical)$ shown is obtained from constant
mobility simulations using Prophet).
The main features of this comparison are:

\noindent
{\bf (a)} Subthreshold region: The slope $d[log(I_d)]/dV_g$ is smaller
in the quantum case when compared to Medici (Fig.
\ref{fig:Id1vsVg_quantum_medici}). Further, the current resulting from
the simple intuitive expression
\begin{eqnarray}
I = I_{q0} \; e^{\frac{-E_{r1}}{kT}} \label{eq:q-curr}
\end{eqnarray}
matches the quantum result quite accurately. $I_{q0}$ is a prefactor
chosen to reproduce the current at $V_g=0$ in Fig.
\ref{fig:Id1vsVg_quantum_medici}. This match is rationalized by noting
that for the values of gate biases considered, $E_{r1}$ is well above
the source Fermi energy and $E_{r2}$ is many kT (thermal energy) above
$E_{r1}$. The difference in slope between the classical and quantum
results can be understood from the slower variation of $E_{r1}$ in
comparison to $E_b(classical)$ as a function of $V_g$ (Fig.
\ref{fig:EbERvsVg}). We also find that the decrease of $E_{r1}$
with increases in gate voltage is slower than the barrier height
determined from the quantum potential profiles. This arises because
(neglecting 2D effects) $E_{r1}$ is determined by a triangular well
(whose apex is the conduction band) that becomes progressively narrower
with increase in gate voltage.

\noindent
{\bf (b)} Large gate biases: The drain current and slope
$d[log(I_d)]/dV_g$ are larger in the quantum case. The higher
$dI_d/dV_g$ at large gate voltages in the quantum case can be
understood from the fact that $E_{r1}$ is above the Fermi level while
$E_b(classical)$ is below, at $V_g=1$V (the quantum current is
proportional to exp($-(E_{r1}-E_F)/kT)$). The mobility model assumed
in the classical case also plays a role in determining the slope.

\subsection{$I_d$ versus $V_d$}
\label{subsection:IdvsVd}

The values of $dI_d/dV_d$ and drive current are important in MOSFET
applications because they determine switching speeds.~ \cite{Thompson}
Figure \ref{fig:IdvsVd_1B_1BP_medici} compares the drain current versus
drain voltage for $V_g=0$ and $V_g=1V$. The drive current ($V_g=1V$)
calculated using Q1 with the polysilicon region treated in the flat
band and q-poly approximations is more than 100\% and 200\% larger than
the results in reference \onlinecite{MIT25inputdeck}. $dI_d/dV_d$ in the linear region
is up to three times larger in Q1. The subthreshold drain current
is smaller in Q1. We however expect that with decreasing channel
length, the sub threshold $I_d$ will become larger than the Medici
results due to quantum mechanical tunneling.~\cite{Naveh00}

\subsection{Isotropic versus anisotropic effective mass}
\label{subsection:effecmass}

The primary influence of anisotropic effective mass is to influence the
energy of the subbands in the inversion layer.
Valleys with the largest effective mass perpendicular to the oxide
(0.98$m^\ast_o$) have subband energies that are smaller than the
isotropic effective mass case. We see from the plot of transmission
versus energy (Fig. \ref{fig:DOSandTvsEbw13B}) that the valleys with
($m_x=0.98 m^\ast_o$, $m_y=m_z=0.19 m^\ast_o$) have resonance levels
that are more than 50meV lower in energy than the isotropic effective
mass case. The corresponding subthreshold current (Fig.
\ref{fig:IdvsVg_1B_1BP_3B_3BP}) is a few hundred percent larger than
the value obtained from the isotropic effective mass case. This follows
by noting that the subthreshold current depends on $exp(-E_{r1}/kT)$.
The drive current (Fig. \ref{fig:IdvsVg_1B_1BP_3B_3BP}) from the
anisotropic effective mass case is more than twenty five percent larger
than the isotropic
effective mass case. Note that for large gate voltages the dependence
of current on $E_{r1}$ is sub exponential. We are not aware of any
calculations that compare the relative importance of the current
carrying capacity of electrons in the three inequivalent valleys. We
find that the valley with the largest $m_x$ (=0.98$m^\ast_o$) carries
89.22 \% and 79.77 \% of the current at $V_g$ equal to 0 and 1V
respectively ($V_d=1$V). Thus all three valleys  are necessary for
an accurate calculation of the ballistic current.


\subsection{Gate leakage current}
\label{subsection:gateleakage}

A major problem in MOSFETs with ultra thin oxides is that tunneling
from gate to
drain will determine the OFF current. The gate leakage current versus
y is plotted for the MIT25 device in Fig. \ref{fig:IgvsYallpoly}.
Note that while we use a value of $3.0$ for the dielectric
constant of SiO$_2$, a value of 3.9 does not change the qualitative
conclusions.
At $V_g=0$V and $V_d=1$V, the main path for leakage current is from
the polysilicon gate contact on top of the oxide to the highly doped
(n$^+$) regions associated with the drain (Source Drain Extension,
SDE) as shown in Fig. \ref{fig:IgvsYallpoly} (a).
At non zero $V_g$, there is also an appreciable tunneling from
the highly doped n$^+$ regions near the source to the polysilicon
region on top of the gate (Fig. \ref{fig:IgvsYallpoly} (a)). For
$t_{ox}=1.5$ nm, gate tunneling increases the OFF current by about two
orders of magnitude, and for smaller oxide thicknesses, the gate
leakage current is significantly larger.

We propose that the gate leakage current can be reduced by a factor of
10-100 without significantly compromising the drive current. The drive
current in these ultra small MOSFETs is primarily determined by the
source injection barrier, or more correctly
as discussed earlier by the resonant level at the source injection
barrier. So any changes that result in a reduction of the gate leakage
current should not significantly alter the location of the resonant
level at the source injection barrier (and hence the drive current).
Two methods (without regard to fabrication issues) that help in this
direction are discussed below:

(i) {\it Shorter or asymmetric polysilicon gate region}:
We propose that the gate leakage current can be significantly reduced
by using shorter gate lengths. The main feature of the shorter gate
lengths is a small overlap between the polysilicon gate and the n$^+$
region near the drain. This is
pictorially represented in Figs. \ref{fig:oxide} (a) and (b) with
'long' and 'short' gate lengths. To simulate the long and short gate
lengths, we consider the doping profile of MIT25 with $L_g =$ 25 nm and
50 nm (gate length in reference \onlinecite{MIT25inputdeck}). The OFF current and
gate leakage current are plotted in Fig. \ref{fig:Idvstox}. We see that
the gate leakage current reduces by more than an order of magnitude, and
the drive current is within two percent of the $L_g =$ 50 nm case, as
desired (see inset of Fig. \ref{fig:Idvstox}). The spatial profile of
gate leakage current for $L_g =$ 25 nm is shown in Fig.
\ref{fig:IgvsYallpoly} (b). Though the gate leakage current reduces
significantly, a drawback of this scheme is the requirement for very
short (approximately equal to the distance between highly doped region
near source and drain) polysilicon gate lengths. A polysilicon gate
placed asymmetrically with respect to y=0 such that its overlap with
the n$^+$ regions near the drain is small, will also serve to
reduce the OFF current without compromising the drive current.

(ii) {\it Graded oxide}:
The second proposal is to use a graded oxide, which is thinner close to
the source end and thicker close to the drain end (Fig. \ref{fig:oxide}
(c)). The thinner oxide near the source is not
expected to alter the source injection barrier significantly,
while the tunneling rate from gate to drain will be significantly
smaller because of the thicker oxide in the drain-gate overlap region.
We consider an oxide that is 1.5 nm thick for y $<$ +10 nm and 2.5 nm
for y $>$ 11 nm, with the thickness varying linearly in between. The
polysilicon gate lengths is 50 nm. Comparison of this device to the
original MIT25 with an uniform oxide and $L_g=50$ nm show that while
the gate leakage current decreases by one order of magnitude, the drive
current decreases by only 30 \%. Further
optimization of this device structure could yield a larger drive
current, while keeping the gate leakage current small.

\section{Concluding Remarks}

A modeling framework and computer code to calculate properties of
ballistic
MOSFETs with open boundaries at the source, drain and gate contacts
have been developed. This includes an algorithm to compute the electron
density using the NEGF equations that avoids solving for the entire
$G^r$ matrix even in the presence of non zero self energies throughout
the device. Note that the simulations presented are
2D in nature and also involve self-consistency. As a result, they were
numerically intensive and were typically performed on sixteen to sixty
four processors of an SGI Origin machine.

The main results of this study are:

(a) Polysilicon gate depletion causes the conduction band close to the
oxide interface to bend in a manner opposite to the semi-classical case
(Fig. 2). This causes a substantial shift in the location of the
conduction band bottom in the channel, which gives rise to drain
currents that are different from the semiclassical case by one to two
orders of
magnitude. Performing quantum mechanical calculations with a flat
polysilicon region, and then shifting the gate voltage axis (in
$I_d$ versus $V_g$) by the quantum mechanical built-in voltage shown
in Fig. 2 results in an order of magnitude better agreement with
results from a quantum mechanical treatment of the polysilicon region.
This built-in voltage can simply be determined
by 1D simulations or an analytical expression. In reality, treatment
of discrete dopants in the polysilicon region will give rise to
results that are in between the 'flat band' and 'q-poly' cases presented
in the paper.

A quantum mechanical treatment of the polysilicon gate region
results in an OFF current ($V_g=0$ V and $V_d=1$ V) that is more than
35 times larger than the OFF current from a flat band treatment of
polysilicon region and published results~\cite{MIT25inputdeck} based
on a sophisticated semiclassical simulator.

\noindent
(b) Resonant levels in the channel the from source to drain increase
the effective source injection barrier for ballistic electrons.
Further, even in the ballistic limit the transmission versus energy
reaches integer values over an energy range that could be many times
the thermal energy. Knowledge of the detailed shape of transmission
versus energy is important to accurately determine the ballistic
current. The precise shape of these transmission steps depends on
the details of the channel to source and drain overlap regions and
the resulting 2D potential profile.  Assuming a sharp step-like
increases in the total transmission is incorrect.

The slope $dI_d/dV_d$, whose importance was emphasized 
in reference \onlinecite{Thompson} and the drive current (at $V_g=1$V) are about 300\%
larger than reported in reference \onlinecite{MIT25inputdeck}. Further, inclusion of
anisotropic effective mass in our calculation makes the quantum results
deviate further from the semiclassical results as shown in Fig.
\ref{fig:IdvsVg_1B_1BP_3B_3BP}.

\noindent
(c) Tunneling of charge across the gate oxide can put a limit on the
OFF current. Models of the tunnel current for thin oxide MOSFETs are
important. We model the gate leakage current in two dimensions and show
that significant reduction in the OFF current is possible without
altering the drive current significantly. This is accomplished by
changing either the gate length (Fig. \ref{fig:oxide} b) or by
introducing a graded oxide (Figs. \ref{fig:oxide} (c)).

\noindent
(d) Quasi-ballistic flow of electrons causes the slope of
$d[log(I_d)]/dV_g$ to be larger than the values obtained from
drift-diffusion methods using field dependent mobility models.

This paper dealt with the modeling of steady state properties of
nanoscale MOSFETs in the ballistic limit. Future work in
quantum mechanical simulation of MOSFETs that are of importance
include:
(i) treatment of scattering mechanisms such as interface roughness
and electron-phonon scattering,~\cite{Ando79,Vasileska97}
(ii) treatment of discrete impurity dopants,~\cite{Asenov00,Frank00}
(iii) switching behavior of MOSFETs / time-dependent 
simulation~\cite{Jauho94,Anantram95} and
(iv) noise characteristics of nano transistors.

\section{Acknowledgements}

We would like to thank: Gerhard Klimeck (JPL) and Mark Lundstrom
(Purdue University) for their interest in our work and for their
useful comments, and Mark Lundstrom and Kent Smith (Bell Laboratory)
for recommending a procedure for faster convergence of Poisson's and
NEGF equations.  The calculations were performed on an SGI Origin 2000
machine located at
NASA Advanced Supercomputing (NAS) Division, whom we acknowledge. We
thank Bron C. Nelson of NAS/SGI for resolving some high performance
computing related questions on the SGI Origin machine and Prabhakar
Shatdarsham for his valuable help with Matlab. We thank Supriyo
Bandyopadyay (UNL - University of Nebraska, Lincoln) and Meyya
Meyyappan (NASA Ames Research Center) for arranging this collaborative
effort between NASA Ames Research Center and UNL.

\pagebreak

\section{Appendix A}

\noindent
{\it Derivation of Eqs. (\ref{eq:g<L2}) and (\ref{eq:g<L})}:

Using Dyson's equation for $G^<$ [Eq. (\ref{eq:dysonG<1})],
we obtain
\begin{eqnarray}
g^{<Lq+1}_{q+1,q+1} =
  g^{r0}_{q+1,q+1} A_{q+1,q}^\dagger
                             g^{<Lq+1}_{q,q+1}
+ g^{r0}_{q+1,q+1} \Sigma^<_{q+1,q+1}
                           g^{aLq+1}_{q+1,q+1}
+ g^{r0}_{q+1,q+1} \Sigma^<_{q+1,q}
       g^{aLq+1}_{q,q+1} \mbox{ .}
                                        \label{eq:temp2}
\end{eqnarray}
Using Eq. (\ref{eq:dysonG<3}), $g^{<Lq+1}_{q,q+1}$ can be
expressed in terms of $g^{<Lq+1}_{q+1,q+1}$, the quantity
we are solving for and known Green's functions as
\begin{eqnarray}
g^{<Lq+1}_{q,q+1} = g^{<0}_{q,q+1}
+ g^{<0}_{q,q+1} A_{q+1,q}^\dagger
  g^{aLq+1}_{q,q+1}
+ g^{<Lq}_{q,q} A_{q,q+1}^\dagger
                       g^{aLq+1}_{q+1,q+1}
+ g^{rLq}_{q,q} A_{q,q+1}
                       g^{<Lq+1}_{q+1,q+1}
\label{eq:temp3} \mbox{ .}
\end{eqnarray}
Substituting Eq. (\ref{eq:temp3}) in Eq. (\ref{eq:temp2}),
we obtain
\begin{eqnarray}
&&\left[I-g^{r0}_{q+1,q+1} A_{q+1,q}
g^{rLq}_{q,q} A_{q,q+1}\right]
g^{<Lq+1}_{q+1,q+1} = \nonumber \\
&&
\;\;\;\;\;\;\;\;\;\;\;\;\;\;\;\;\;\;\;\;\;\;\;\;\;\;\;\;\;\;\;\;\;\;
g^{r0}_{q+1,q+1} \Sigma^<_{q+1,q+1}
                           g^{aLq+1}_{q+1,q+1}
+ g^{r0}_{q+1,q+1} \Sigma^<_{q+1,q}
       g^{aLq+1}_{q,q+1}
\nonumber \\
&&
\;\;\;\;\;\;\;\;\;\;\;\;\;\;\;\;\;\;\;\;\;\;\;\;\;
+ g^{r0}_{q+1,q+1} A_{q+1,q}
                               \left[ g^{<0}_{q,q+1}
+ g^{<0}_{q,q+1} A_{q+1,q}^\dagger
                              g^{aLq+1}_{q,q+1}
+ g^{<Lq}_{q,q} A_{q,q+1}^\dagger
                    g^{aLq+1}_{q+1,q+1} \right]
\mbox{ .}             \label{eq:gtemp1}
\end{eqnarray}
Using Eq. (\ref{eq:dysongrL2}) and $g^{<0}_{q,q+1} = g^{rLq}_{q,q}
\Sigma^<_{q,q+1} g^{a0}_{q+1,q+1}$, which follows from
Eq. (\ref{eq:G<0}), we obtain
\begin{eqnarray}
g^{<Lq+1}_{q+1,q+1}
&=&
g^{rLq+1}_{q+1,q+1} \left[\Sigma^{<}_{q+1,q+1}
+ A_{q+1,q} g^{<Lq}_{q,q} A_{q,q+1}^\dagger \right]
                                  g^{aLq+1}_{q+1,q+1}
\nonumber \\
& & \;\;\;\;\;\;\;\;\;\;\;\;\;\;\;\;\;
+
g^{rLq+1}_{q+1,q+1} \Sigma^{<}_{q+1,q}
g^{aLq+1}_{q,q+1}
+
g^{rLq+1}_{q+1,q+1} A_{q+1,q}
g^{rLq}_{q,q} \Sigma^{<}_{q,q+1}
g^{aLq+1}_{q+1,q+1} \mbox{ .}
\label{eq:g<L1}
\end{eqnarray}
Noting that $g^{rLq+1}_{q+1,q}= g^{rLq+1}_{q+1,q+1} A_{q+1,q}
g^{rLq}_{q,q}$, Eq. (\ref{eq:g<L1}) can be written as
\begin{eqnarray}
g^{<Lq+1}_{q+1,q+1}
&=&
  g^{rLq+1}_{q+1,q+1} \left[\Sigma^{<}_{q+1,q+1}
+ \sigma^<_{q+1} \right] g^{aLq+1}_{q+1,q+1}
+ g^{rLq+1}_{q+1,q+1} \Sigma^{<}_{q+1,q} g^{aLq+1}_{q,q+1} \nonumber \\
& & \;\;\;\;\;\;\;\;\;\;\;\;\;\;\;\;\;\;\;\;\;\;\;\;\;\;\;\;\;\;\;\;\;
+ g^{rLq+1}_{q+1,q} \Sigma^{<}_{q,q+1}
          g^{aLq+1}_{q+1,q+1} \mbox{,} \label{eq:g<L2p} \\
    \mbox{where,}
& & \nonumber \\
\sigma^<_{q+1} &=& A_{q+1,q} g^{<Lq}_{q,q} A_{q,q+1}^\dagger
\mbox{ .}              \label{eq:sigmaqp1}
\end{eqnarray}

\pagebreak

\noindent
{\bf Figure Captions:}

\begin{figure}[htbp]
  \begin{center}
    \leavevmode
    \epsfxsize=3.375in
    \epsfbox{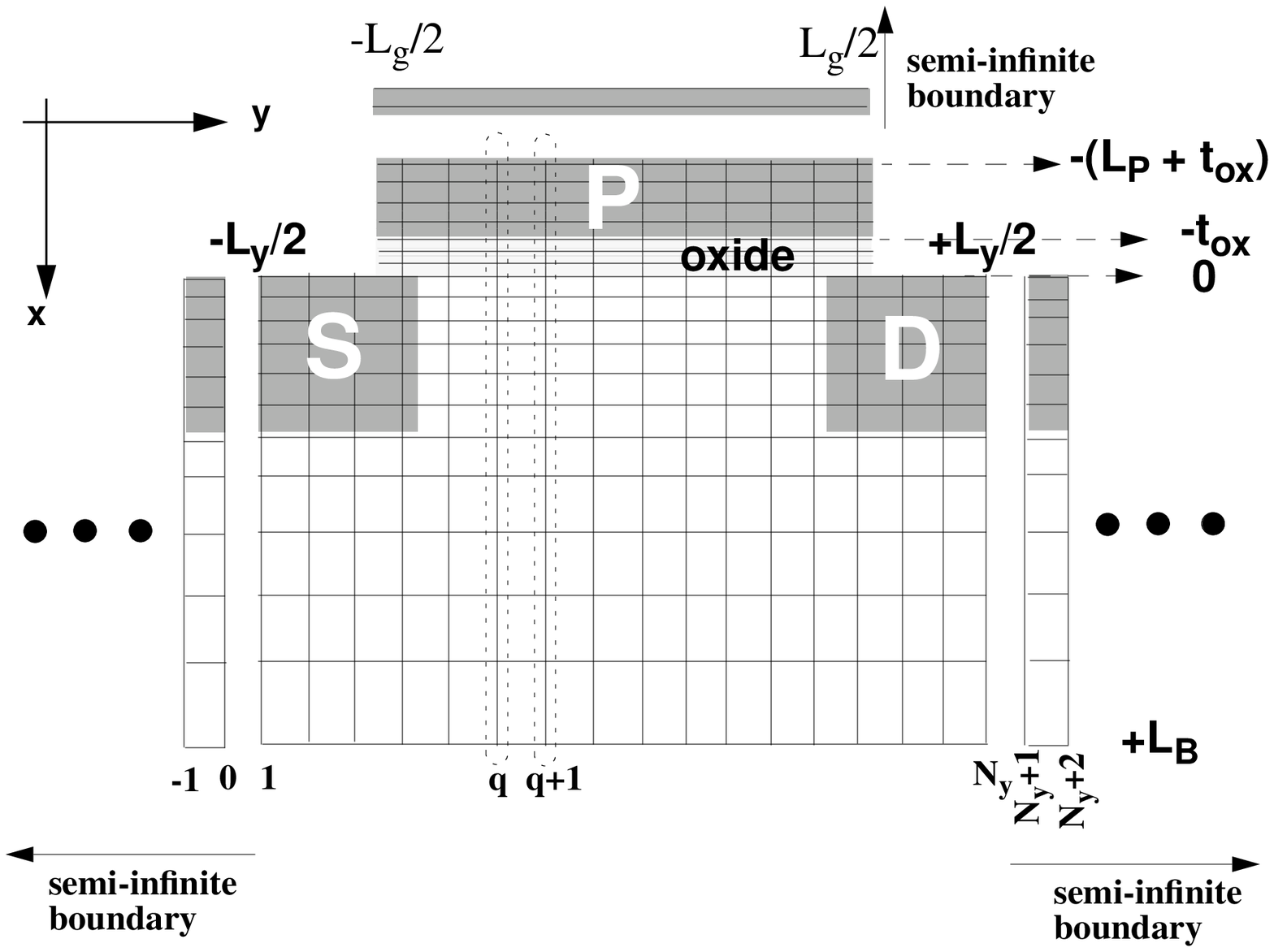}
  \end{center}
  \caption{}
  \label{fig:grid}
\end{figure}

Fig. \ref{fig:grid}: The equations are solved in a 2D non uniform
spatial grid, with semi-infinite boundaries as shown. Each column $q$
comrises the diagonal blocks of Eqs. (\ref{eq:discreteGr2}) and
(\ref{eq:discreteG<2}). The electrostatic potential is held fixed at
the begining of the semi-infinte regions closest to the device.

\begin{figure}[htbp]
  \begin{center}
    \leavevmode
    \epsfxsize=3.375in
    \epsfbox{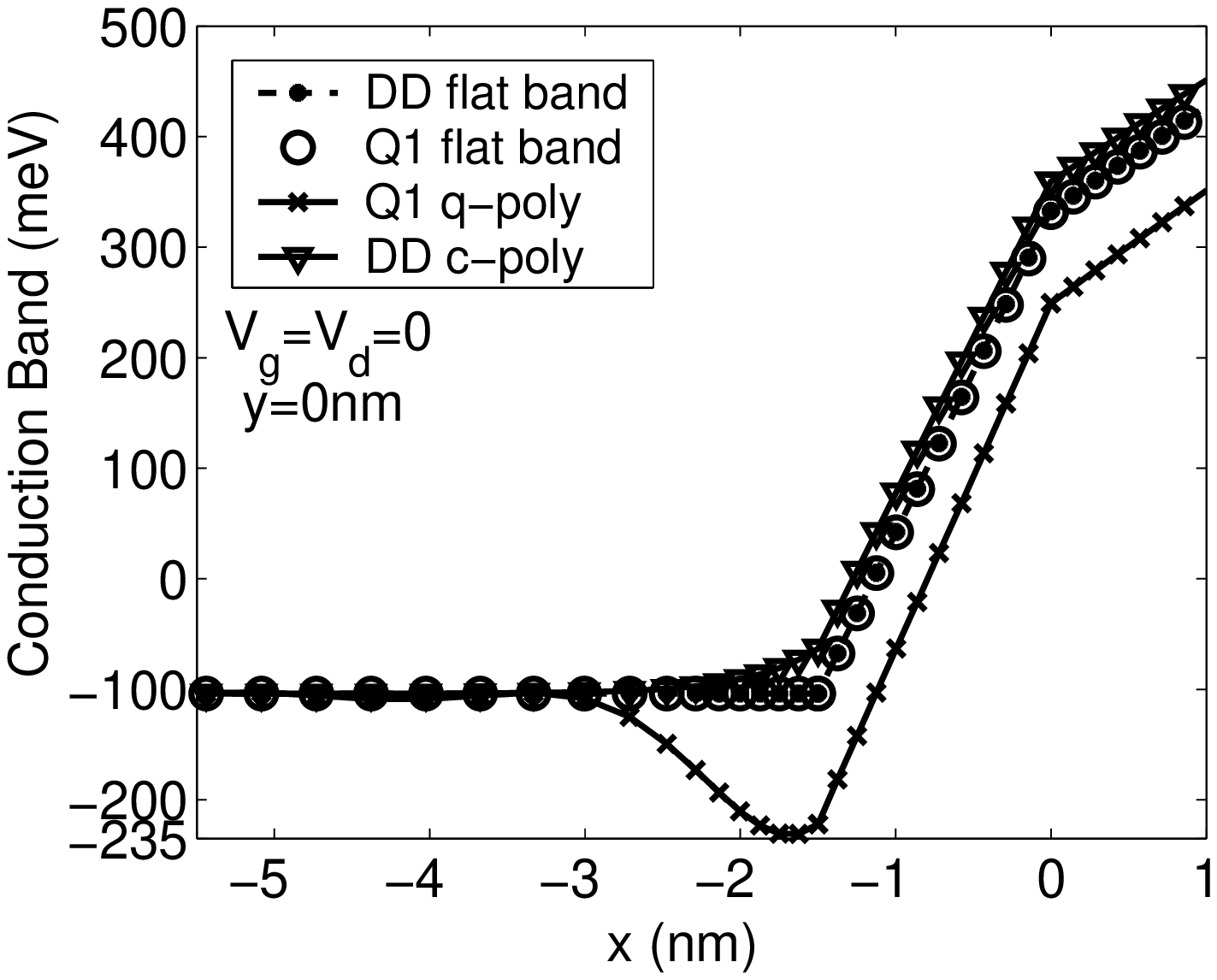}
  \end{center}
  \caption{}
  \label{fig:PolyFlatBand}
\end{figure}

Fig. \ref{fig:PolyFlatBand}:
Potential profile at the y=0 slice of MIT25, calculated by
four different methods. Note the qualitative difference of the
'Q1 q-poly' case due to electron depletion in the gate.

\begin{figure}[htbp]
  \begin{center}
    \leavevmode
    \epsfxsize=3.375in
    \epsfbox{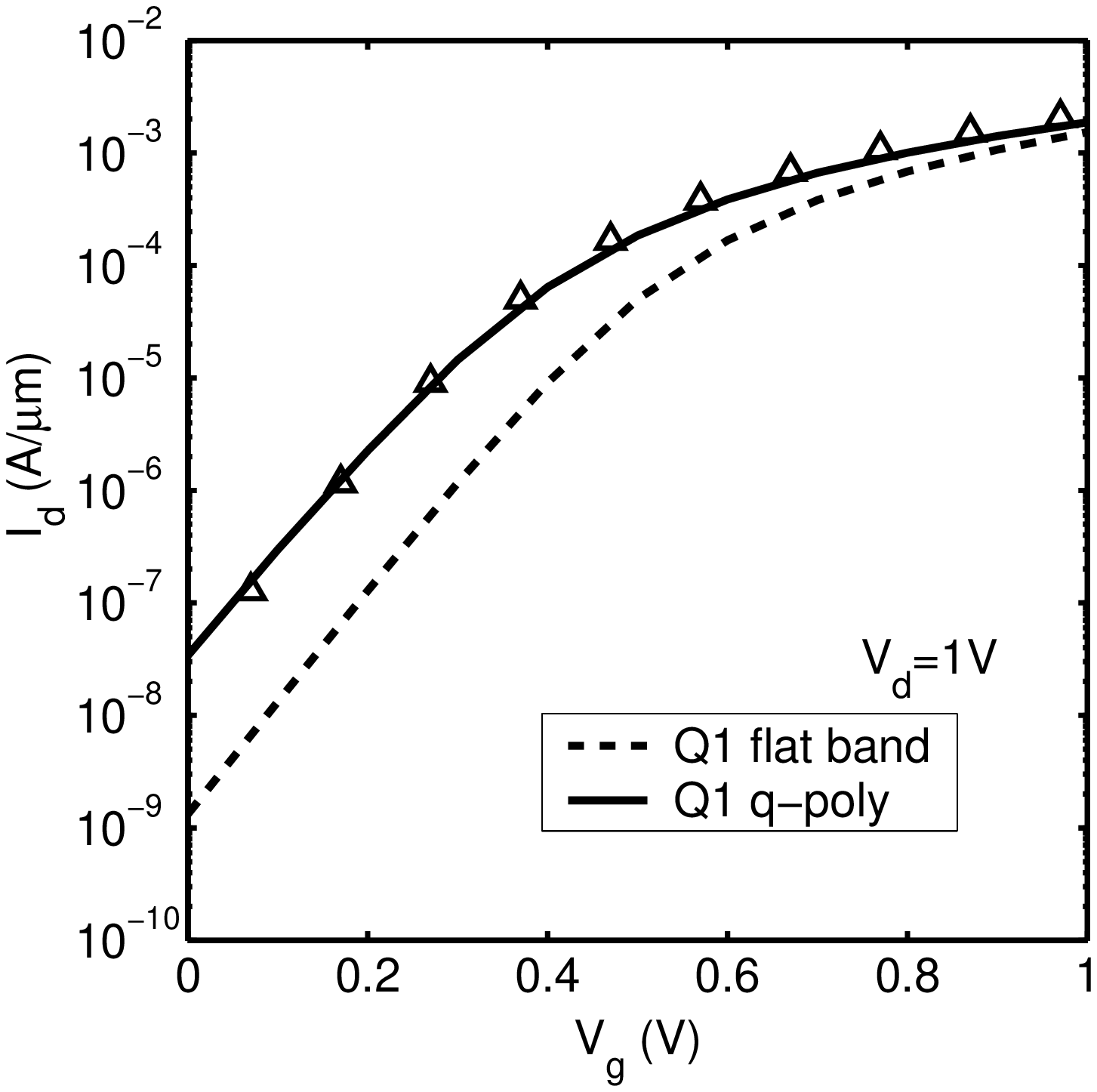}
  \end{center}
  \caption{}
  \label{fig:IdvsVg_1B_1BP}
\end{figure}

Fig. \ref{fig:IdvsVg_1B_1BP}:
Drain current versus gate voltage for $V_d=1$ V. Quantum mechanical
treatment of the polysilicon gate (Q1 q-poly)  results in much higher
current.

\begin{figure}[htbp]
  \begin{center}
    \leavevmode
    \epsfxsize=3.375in
    \epsfbox{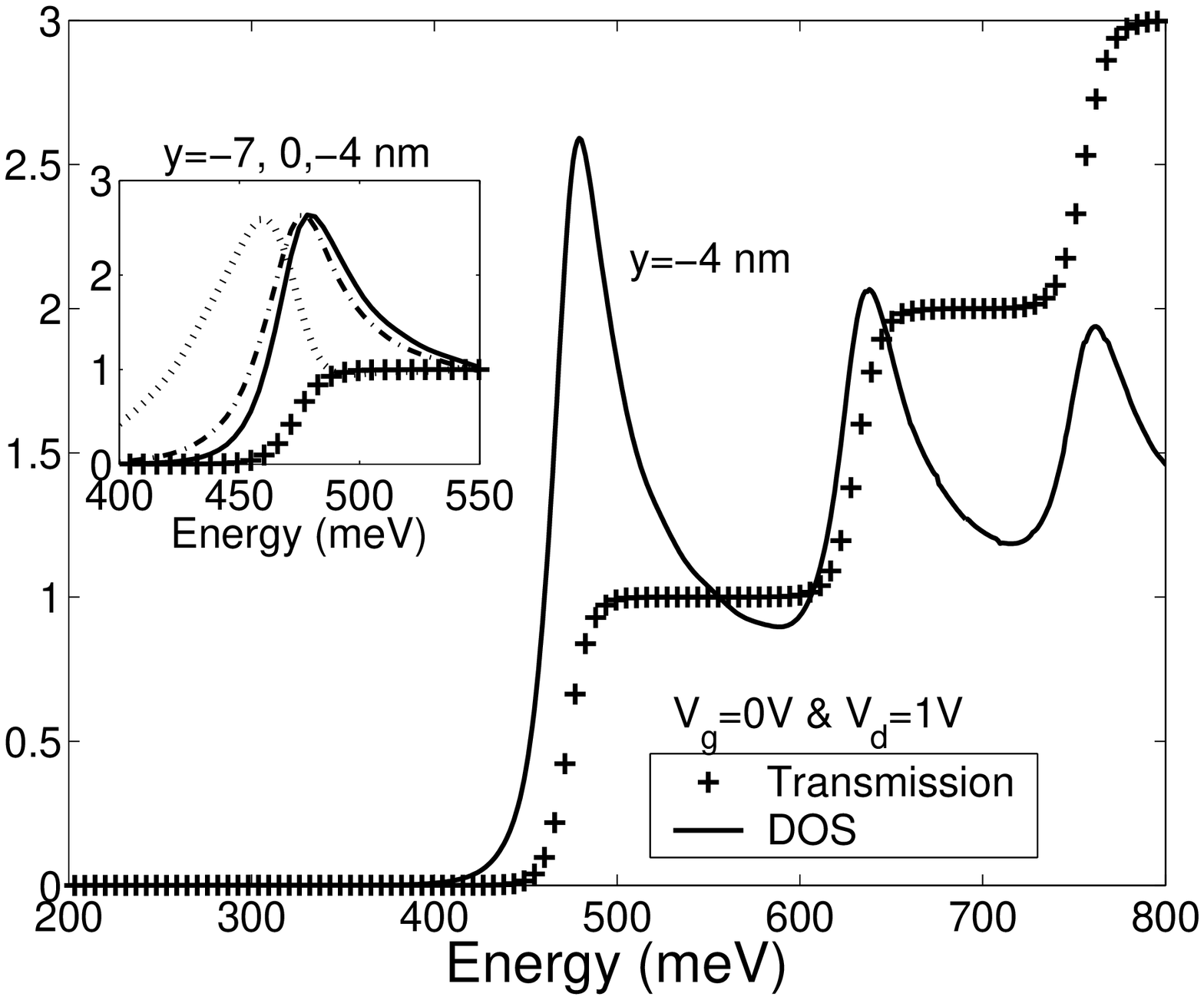}
  \end{center}
  \caption{}
  \label{fig:DOSandTvsEbw}
\end{figure}

Fig. \ref{fig:DOSandTvsEbw}:
Transmission (+) and density of states (DOS) versus energy at a
spatial location close to the source injection barrier, at $V_g=0$V
and $V_d=1$V. The peaks in the density of states represent the
resonant levels in the channel.
Inset: DOS at three different y-locations and the total transmission.
The points y = -7 and 0 nm are to the left and right of the location
where the source injection barrier is largest (close to y = -4 nm).

\begin{figure}[htbp]
  \begin{center}
    \leavevmode
    \epsfxsize=3.375in
    \epsfbox{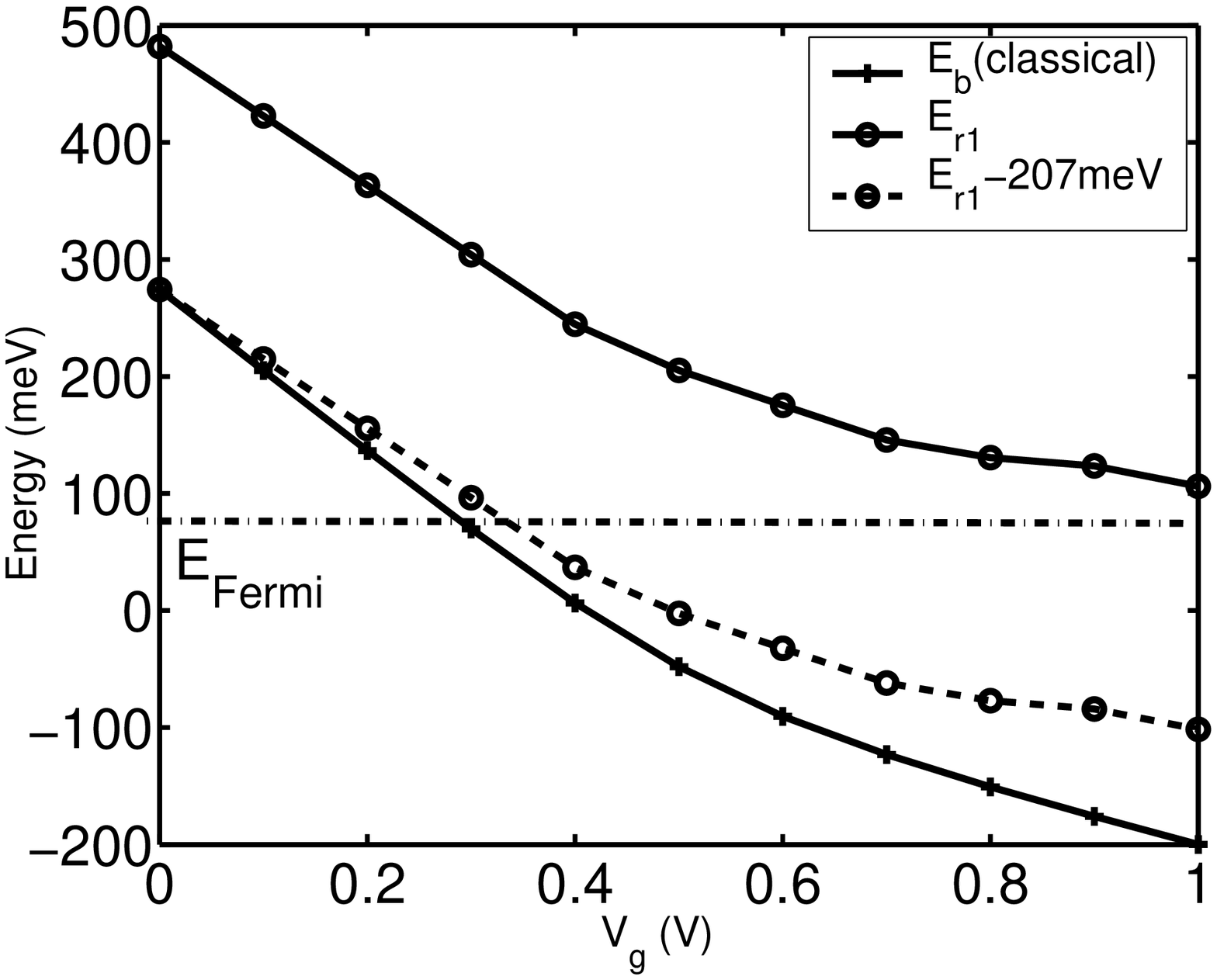}
  \end{center}
  \caption{}
  \label{fig:EbERvsVg}
\end{figure}

Fig. \ref{fig:EbERvsVg}:
Location of the first resonant level ($E_{r1}$) versus gate voltage
and the classical source injection barrier ($E_b(classical)$). Note
that $E_{r1}$ decreases slower than $E_b(classical)$ with gate
voltage due to narrowing of channel potential well.

\begin{figure}[htbp]
  \begin{center}
    \leavevmode
    \epsfxsize=3.375in
    \epsfbox{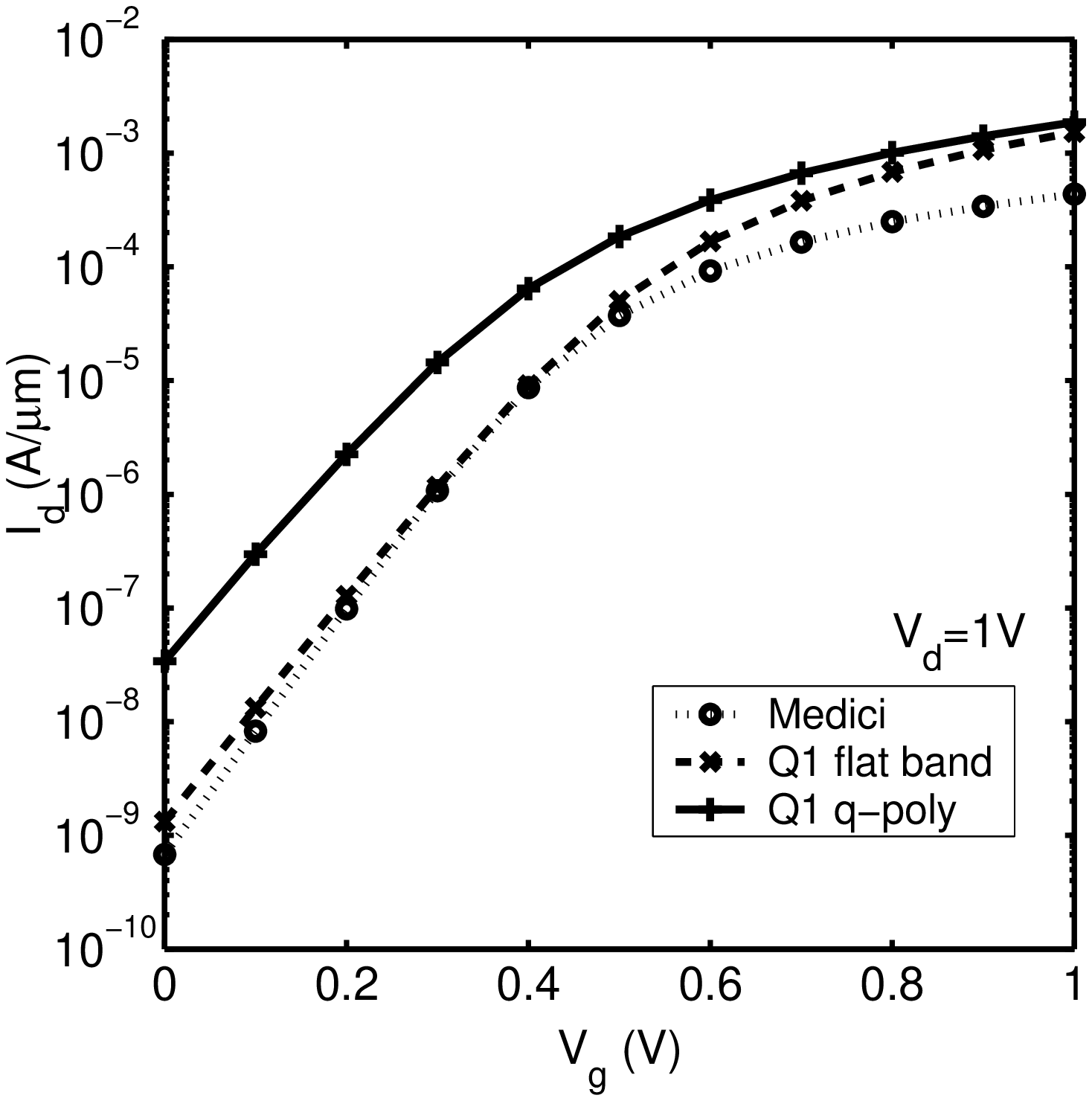}
  \end{center}
  \caption{}
  \label{fig:Id1vsVg_quantum_medici}
\end{figure}

Fig. \ref{fig:Id1vsVg_quantum_medici}:
Plot of drain current versus gate voltage from the quantum mechanical
calculations and Medici, at $V_d=1$V. At small gate voltages, the
drain current from Medici~\cite{MIT25inputdeck} are comparable to the
'Q1 flat band' results. The drain current from 'Q1 q-poly' is however
significantly different at large gate voltages.

\begin{figure}[htbp]
  \begin{center}
    \leavevmode
    \epsfxsize=3.375in
    \epsfbox{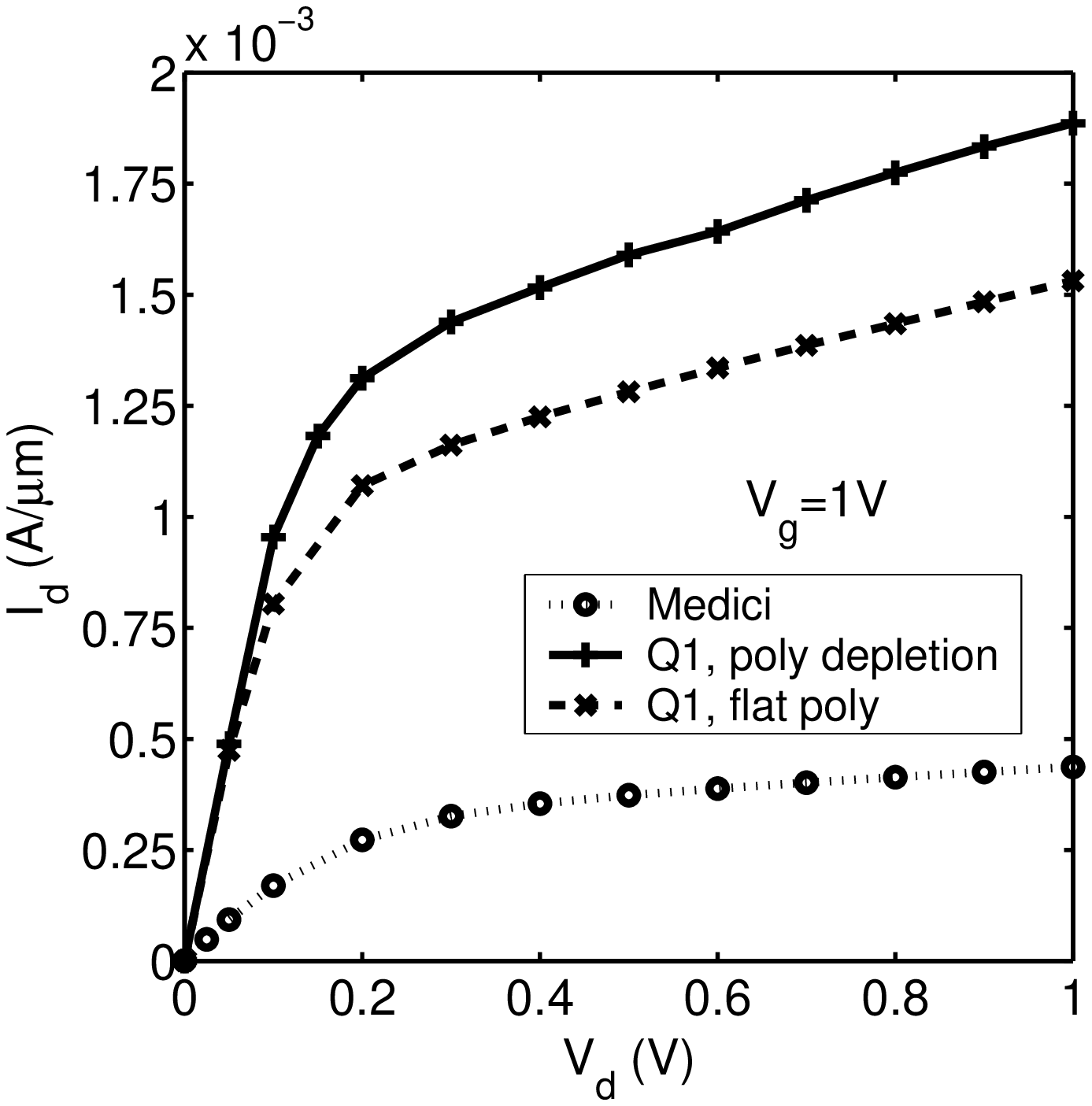}
  \end{center}
  \caption{}
  \label{fig:IdvsVd_1B_1BP_medici}
\end{figure}

Fig. \ref{fig:IdvsVd_1B_1BP_medici}:
Plot of drain current versus drain voltage ($V_d$) from the quantum
mechanical calculations and Medici, at $V_g=1$V. Note the large
difference in drive current and $dI_d/dV_d$ between 
Medici,~\cite{MIT25inputdeck} 'Q1 flat band' and 'Q1 q-poly'.

\begin{figure}[htbp]
  \begin{center}
    \leavevmode
    \epsfxsize=3.375in
    \epsfbox{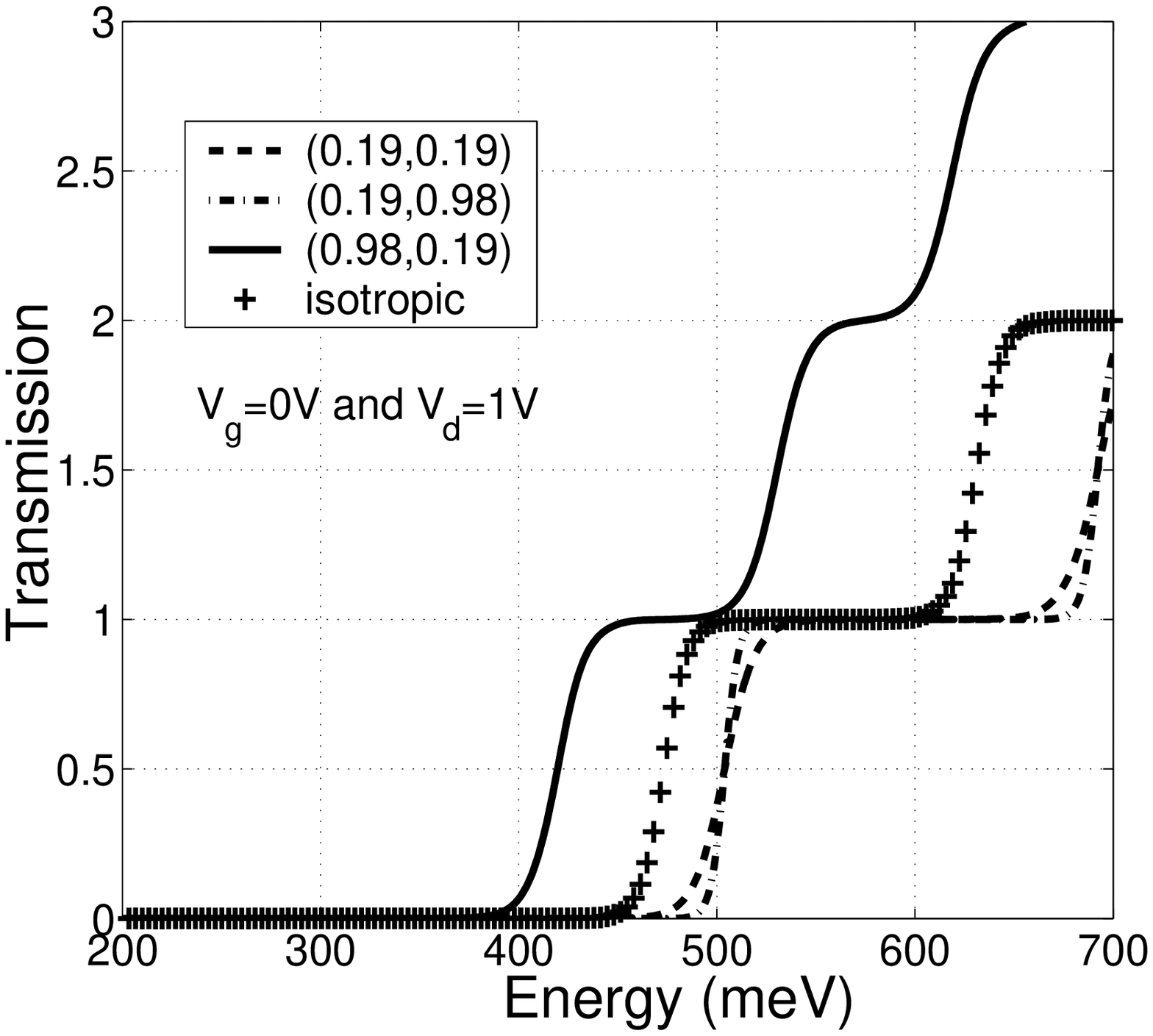}
  \end{center}
  \caption{}
  \label{fig:DOSandTvsEbw13B}
\end{figure}

Fig. \ref{fig:DOSandTvsEbw13B}:
Same as Fig. \ref{fig:DOSandTvsEbw} but the anisotropic effective mass
case is included. Note that the valley with the largest mass in the
x-direction has subband energies that are about 50 meV smaller than
the isotropic effective mass case even at $V_g=0$.

\begin{figure}[htbp]
  \begin{center}
    \leavevmode
    \epsfxsize=3.375in
    \epsfbox{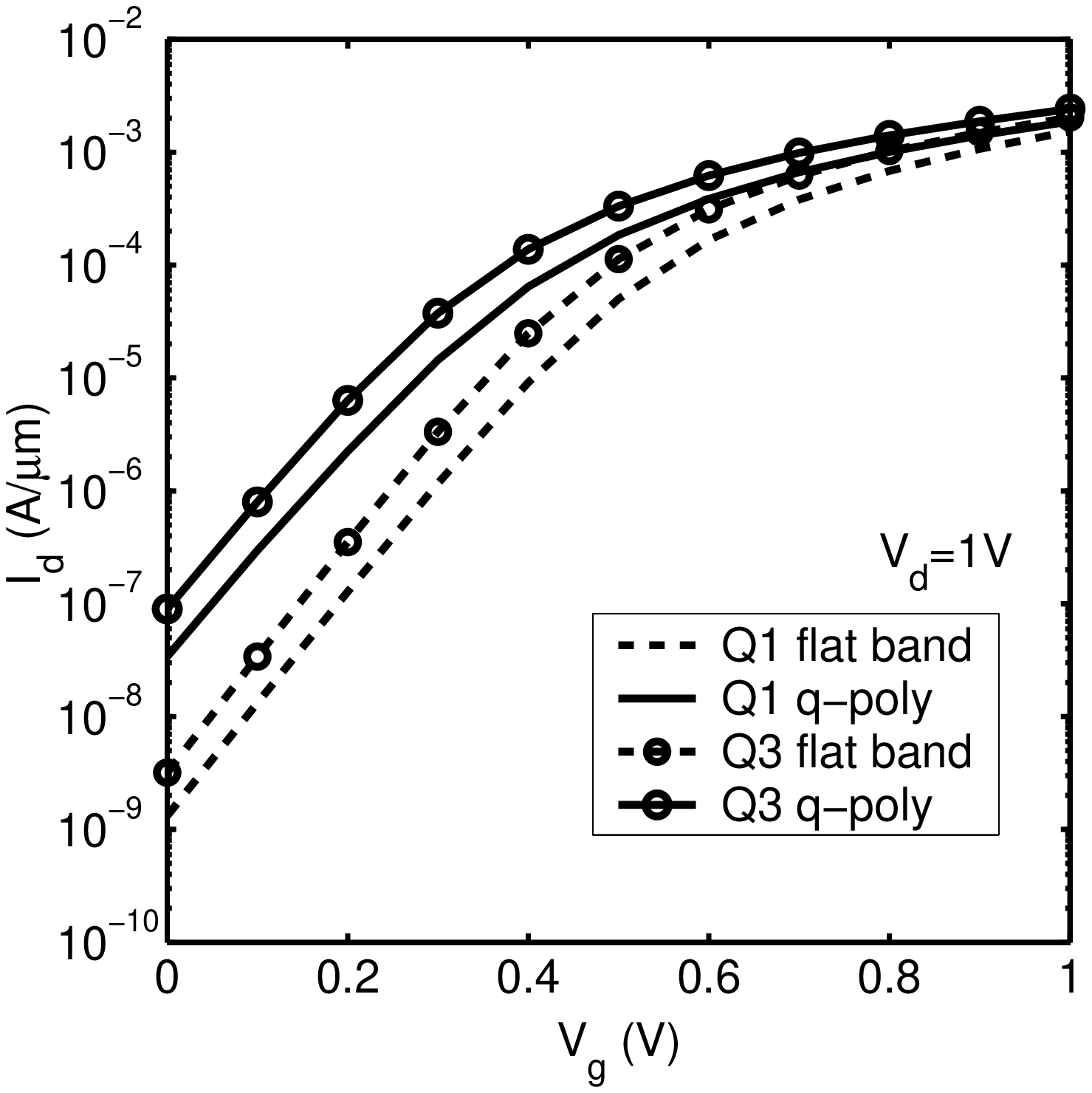}
  \end{center}
  \caption{}
  \label{fig:IdvsVg_1B_1BP_3B_3BP}
\end{figure}

Fig. \ref{fig:IdvsVg_1B_1BP_3B_3BP}:
Plot of drain current versus gate voltage for the isotropic and
anisotropic effective mass cases, at $V_d=1$V. The much higher current
in the anisotropic effective mass case (Q3) is due to the lower
suband energy shown in Fig. \ref{fig:DOSandTvsEbw13B}.

\begin{figure}[htbp]
  \begin{center}
    \leavevmode
    \epsfxsize=3.375in
    \epsfbox{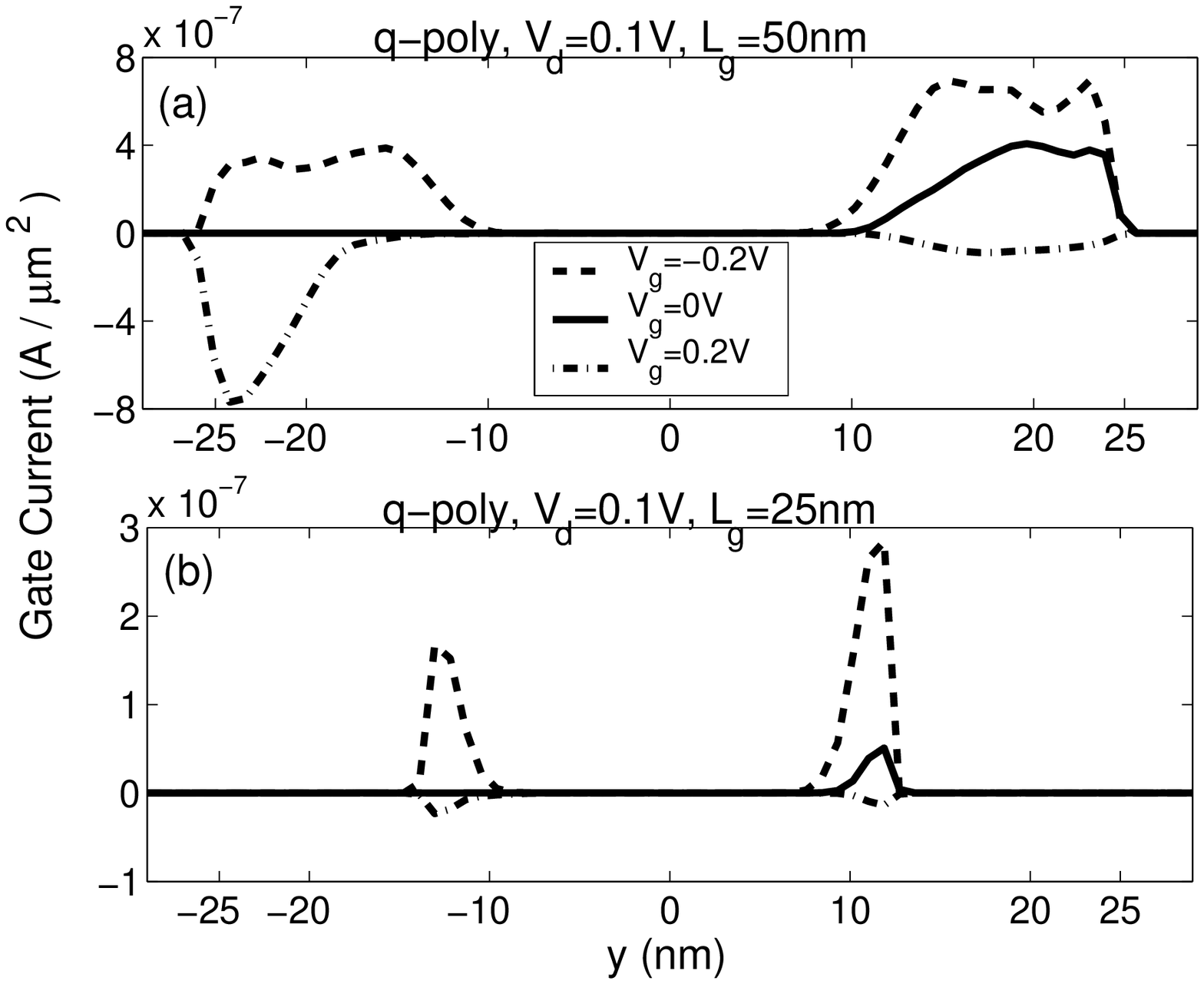}
  \end{center}
  \caption{}
  \label{fig:IgvsYallpoly}
\end{figure}

Fig. \ref{fig:IgvsYallpoly}:
Plot of gate leakage current when the device is OFF ($V_g=0$V) as a
function of the y-direction, from the source to drain, for $L_g$ equal
to (a) 50 nm and (b) 25 nm. Note the significant gate leakage current
in the regions where the high doping in the source and drain overlap
the gate in (a). A shorter gate eliminates a large fraction of the
gate leakage current as shown in (b).

\begin{figure}[htbp]
  \begin{center}
    \leavevmode
    \epsfxsize=3.375in
    \epsfbox{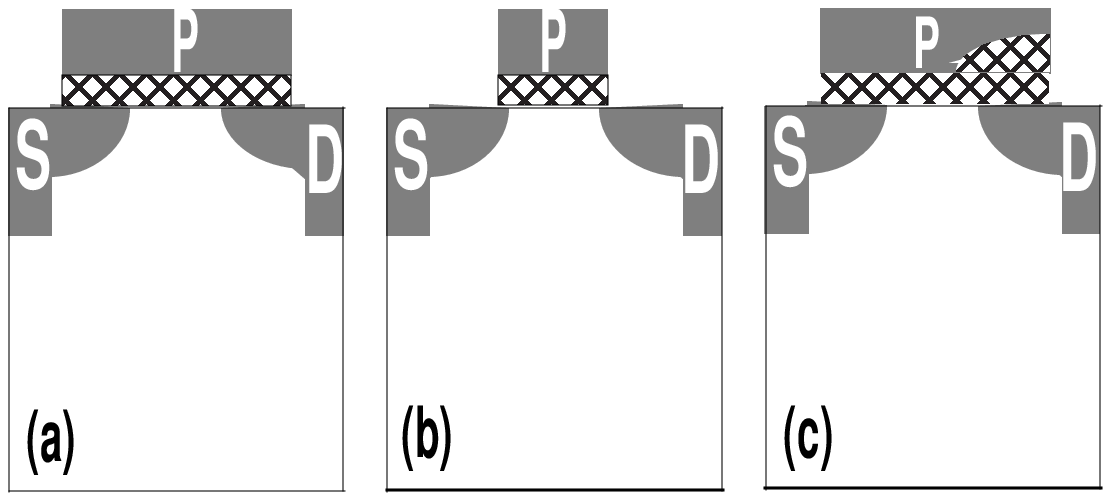}
  \end{center}
  \caption{}
  \label{fig:oxide}
\end{figure}

Fig. \ref{fig:oxide}:
Polysilicon gate and oxide configurations that could reduce the
OFF current ($V_g=0$V) significantly without drastically reducing the
drive current ($V_g=1$V). The hatched marks represent the oxide.

\begin{figure}[htbp]
  \begin{center}
    \leavevmode
    \epsfxsize=3.375in
    \epsfbox{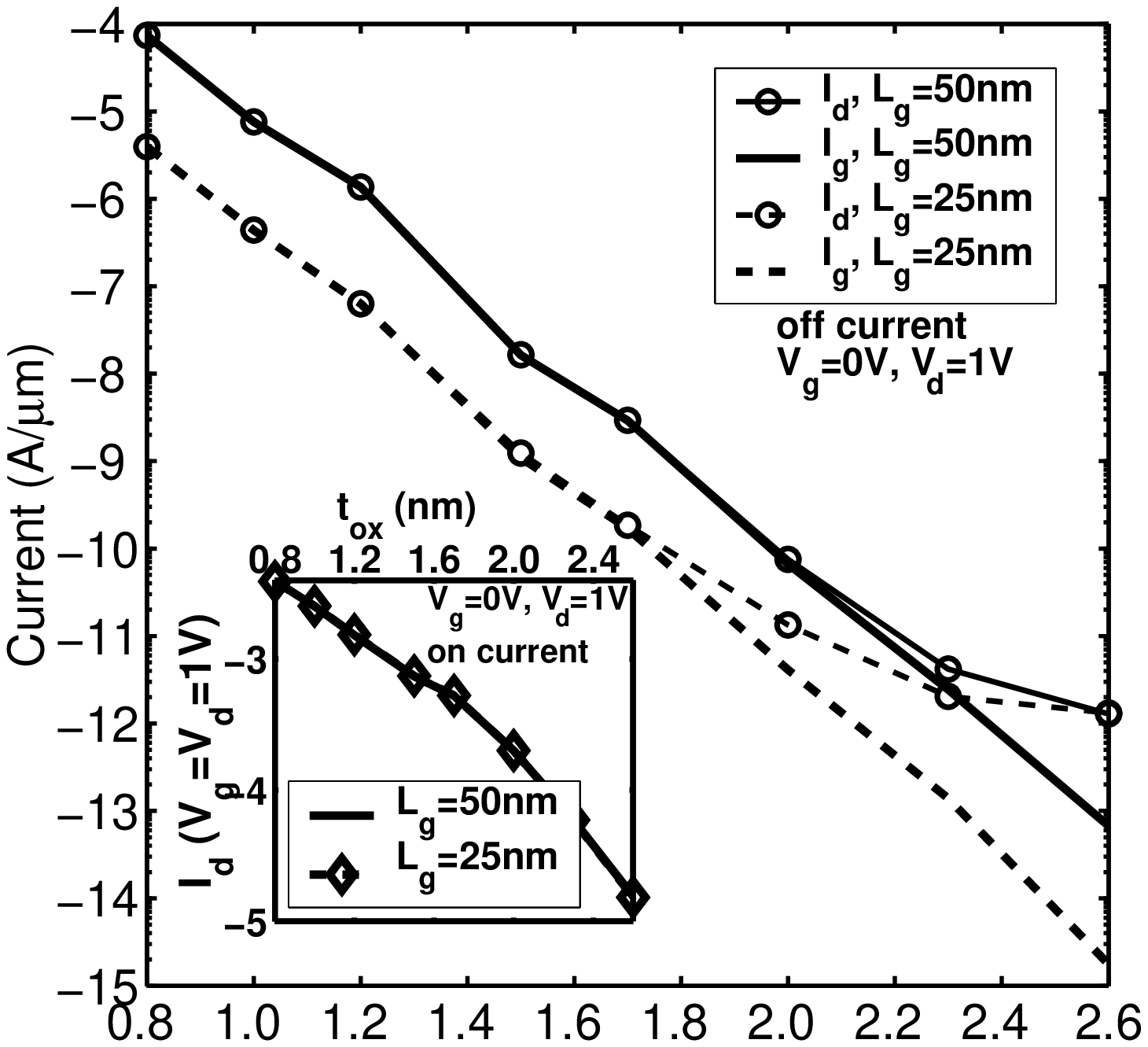}
  \end{center}
  \caption{}
  \label{fig:Idvstox}
\end{figure}

Fig. \ref{fig:Idvstox}:
Plot of drain and gate currents when the device is OFF ($V_g=0$V)
versus oxide thickness for $L_g$ equal to 50 and 25 nm. Inset: Drain
current for the the gate lengths when the device is on ($V_g=1$V). At
the larger values of $t_{ox}$, the gate current ($I_g$) is
significantly smaller than the drain current ($I_d$), meaning that
the drain current is determined by electron injected from the source
to drain. At smaller values of $t_{ox}$, the drain current is
dominated by the gate leakage current as can be seen by comparing
$I_d$ and $I_g$ in this figure. More importantly, note that the
shorter gate length ($L_g=$ 25 nm) gives an order of magnitude
smaller drain current when the device is OFF for the smaller vaues
of $t_{ox}$. The inset shows that the drive current ($V_d=V_g=1$ V)
is however not affected much by the shorter gate length.

\pagebreak

\pagebreak

\pagebreak

\pagebreak

\pagebreak

\pagebreak

\pagebreak

\pagebreak

\pagebreak

\pagebreak

\pagebreak

\pagebreak

\pagebreak

\pagebreak

\begin{figure}[htbp]
  \begin{center}
    \leavevmode
    \epsfxsize=5.375in
    \epsfbox{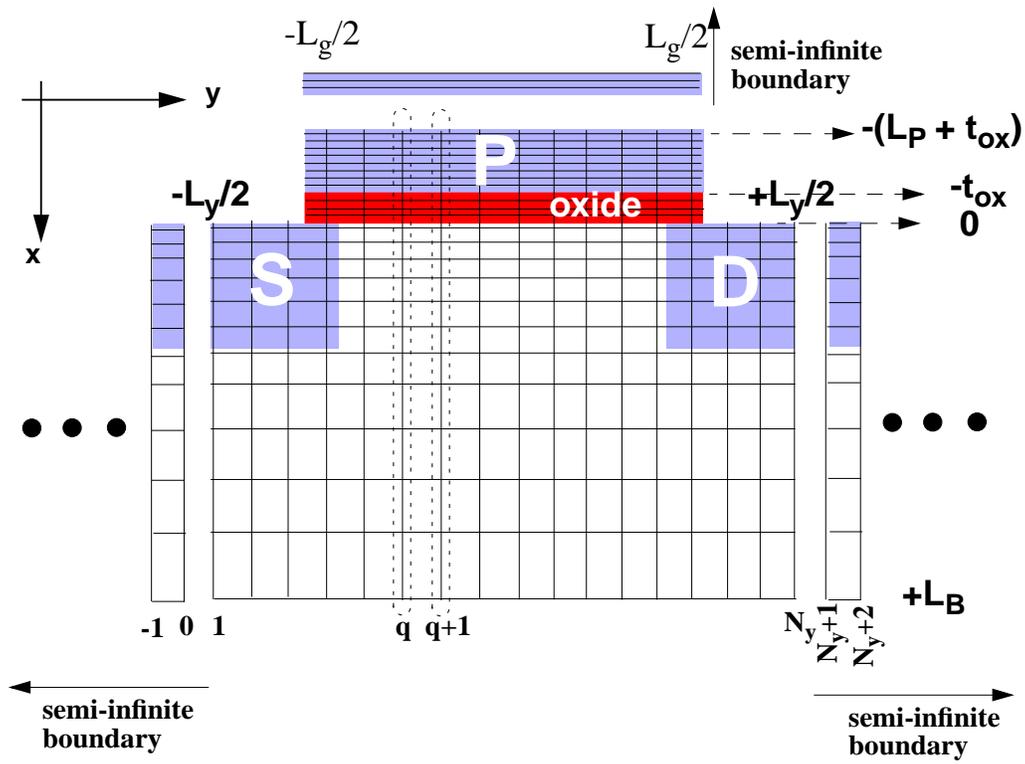}
  \end{center}
  \caption{This figure is a larger version of Fig. 1}
\end{figure}

\end{document}